\documentclass[conference,10pt]{IEEEtran}
\usepackage{amsmath}
\usepackage{amssymb}
\usepackage{amsfonts}
\usepackage{graphicx}
\usepackage{enumerate}
\usepackage{bbm}

\usepackage{mathabx} 
\usepackage{mathtools} 
\usepackage{cite}
\usepackage{subfigure}
\usepackage{url}
\usepackage{array}
\usepackage{verbatim}
\usepackage{comment}
\usepackage{stfloats}

\usepackage{color}
\definecolor{red}{rgb}{1,0,0}
\definecolor{blue}{rgb}{0,0,1}

\newcommand{\expect}[1]{{\mathbb{E}}\left[{#1}\right]}
\newcommand{\E}[1]{\expect{#1}}
\long\def\comment#1{}

\newfont{\bbb}{msbm10 scaled 700}

\newfont{\bb}{msbm10 scaled 1100}

\newcommand{\av}{{\bf a}}
\newcommand{\bv}{{\bf b}}
\newcommand{\cv}{{\bf c}}

\newcommand{\ev}{{\bf e}}
\newcommand{\fv}{{\bf f}}
\newcommand{\gv}{{\bf g}}
\newcommand{\hv}{{\bf h}}

\newcommand{\pv}{{\bf p}}

\newcommand{\rv}{{\bf r}}

\newcommand{\uv}{{\bf u}}
\newcommand{\wv}{{\bf w}}
\newcommand{\vv}{{\bf v}}
\newcommand{\xv}{{\bf x}}
\newcommand{\yv}{{\bf y}}

\newcommand{\zerov}{{\bf 0}}
\newcommand{\onev}{{\bf 1}}

\newcommand{\Am}{{\bf A}}

\newcommand{\Fm}{{\bf F}}

\newcommand{\Hm}{{\bf H}}
\newcommand{\Id}{{\bf I}}

\newcommand{\Pm}{{\bf P}}

\newcommand{\Rm}{{\bf R}}

\newcommand{\Vm}{{\bf V}}

\newcommand{\Cc}{{\cal C}}

\newcommand{\Nc}{{\cal N}}

\newcommand{\Pc}{{\cal P}}

\newcommand{\deltav}{\hbox{\boldmath$\delta$}}

\newcommand{\lambdav}{\hbox{\boldmath$\lambda$}}

\newcommand{\Lambdam}{\hbox{\boldmath$\Lambda$}}

\newcommand{\wsf}{{\sf w}}

\newcommand{\diag}{{\hbox{diag}}}

\newcommand{\var}{{\hbox{var}}}

\newcommand{\eqdef}{\stackrel{\Delta}{=}}
\newcommand{\defines}{{\,\,\stackrel{\scriptscriptstyle \bigtriangleup}{=}\,\,}}

\newcommand{\herm}{{\sf H}}

\newcommand{\transp}{{\sf T}}

\newcommand{\MB}{M}

\newcommand{\MU}{{\tilde{M}}}
\newcommand{\mus}{{\tilde{m}}}
\newcommand{\Fmb}{\Fm}
\newcommand{\Fmu}{{\tilde{\Fm}}}

\newcommand{\fvu}{\tilde{\fv}}

\newcommand{\width}{\wsf}

\newcommand{\Vmb}{\Vm}
\newcommand{\vvb}{\vv}

\newcommand{\infosigv}{\uv}

\newcommand{\infosig}{u}

\newcommand{\avb}{\av}
\newcommand{\avu}{\tilde{\av}}

\newcommand{\thetab}{\theta}
\newcommand{\thetau}{\tilde{\theta}}

\newcommand{\detus}[2]{D_{#1,#2}}

\newcommand{\detushat}[2]{\hat{D}_{#1,#2}}

\newcommand{\detsec}[1]{{\cal D}_{#1}}
\newcommand{\uicode}[1]{{\cal K}_{#1}}

\newcommand{\codemat}[1]{\Pm_{#1}}
\newcommand{\codemath}[1]{\Pm^\herm_{#1}}
\newcommand{\secenv}[2]{\rv_{#1,#2}}

\DeclareMathOperator*{\ORop}{OR}

\newcommand{\presus}[2]{X_{#1,#2}}

\newcommand{\presushat}[2]{\hat{X}_{#1,#2}}

\newcommand{\presec}[1]{{\cal X}_{#1}}
\newcommand{\barlam}[2]{\bar{\lambda}_{#1,#2}}
\newcommand{\barlamv}[2]{\bar{\lambdav}_{#1,#2}}

\newcommand{\barlamb}[3]{\bar{\lambda}_{#1,#2}^{(#3)}}

\newcommand{\presusb}[3]{X_{#1,#2}^{(#3)}}

\newcommand{\HV}{\check{\Hm}} 

\newcommand{\Hvi}{\check{H}} 

\newcommand{\hV}{\check{\hv}}
\newcommand{\spanZ}{\rm span}

\newcommand{\Rmod}[1]{\Rm_k^{\rm #1}}

\def\argmax{\mathop{\rm arg\,max}}

\begin{document}
\title{ 
Directional Training and Fast Sector-based Processing Schemes for mmWave Channels}

\author{\IEEEauthorblockN{Zheda Li$^1$, Nadisanka Rupasinghe $^2$, Ozgun Y. Bursalioglu$^3$, Chenwei Wang$^3$, Haralabos Papadopoulos$^3$, \\Giuseppe Caire$^4$}
\IEEEauthorblockA{$\ ^1$Dept. of EE, University of Southern California, Los Angeles \\
$\ ^2$Dept. of ECE North Carolina State University, Raleigh, NC\\
$\ ^3$Docomo Innovations Inc, Palo Alto, CA\\
$\ ^4$Technische Universit{\"a}t Berlin, Germany\\
zhedali@usc.edu, rprupasi@ncsu.edu, \{obursalioglu, cwang, hpapadopoulos\}@docomoinnovations.com,\, caire@tu-berlin.de}}

\maketitle
\begin{abstract}
We consider a single-cell scenario involving a single base station (BS) with a massive array serving multi-antenna terminals in the downlink of a mmWave channel.  We present a class of multiuser MIMO schemes, which rely on uplink training from the user terminals, and on uplink/downlink channel reciprocity.    The BS employs {\em virtual} sector-based processing according to which, user-channel estimation and data transmission are performed in parallel over non-overlapping angular sectors.  

The uplink training schemes we consider are non-orthogonal, that is, we allow multiple users to transmit pilots on the same pilot dimension (thereby potentially interfering with one another).   Elementary processing allows each sector to determine the  subset of user channels that can be resolved on the sector (effectively pilot contamination free) and, thus, the subset of users that can be served by the sector. This allows resolving multiple users on the same pilot dimension at different sectors, thereby increasing the overall multiplexing gains of the system. 
Our analysis and simulations reveal that, by using appropriately designed directional training beams at the user terminals, the sector-based transmission schemes we present can yield substantial
 spatial multiplexing and ergodic user-rates improvements with respect to their orthogonal-training counterparts.
\end{abstract}



\vspace{-0.2cm}
\section{Introduction}
\label{section:introduction}
\vspace{-0.15cm}


5G standardization efforts and deployments are projected to bring great performance gains with respect to their predecessors in a multitude of  performance metrics, including user and cell throughput, end-to-end delay,  and massive device connectivity. It is widely expected that these 5G requirements will be met by utilizing  a combination of additional resources, including newly available licensed and unlicensed bands, network densification, large antenna arrays, and new PHY/network layer technologies.  To meet the throughput/unit area requirements, for instance, 5G systems  would need to provide much higher spatial multiplexing gains (e.g.,  number of users served simultaneously) than their 4G counterparts. 

Large antenna arrays and massive MIMO are considered as key technologies for 5G and beyond.
It is expected that new generation deployments would have to utilize the cm and mmWave bands where wide chunks of spectrum are readily available.   Note that the spacing of antenna arrays is  proportional to the wavelength, at mmWave, so large arrays can be packed even on small footprints. Such large-size arrays will be critical  in combatting with the harsher propagation characteristics experienced at mmWave. 

Massive MIMO, originally introduced in \cite{marzetta2006, marzetta-massive},  can yield large spectral efficiencies and spatial multiplexing gains  through the use of a large number of antennas at the base stations (BSs).  Large arrays enable focusing the radiated signal power and creating sharp beams to several users simultaneously,  allowing a BS to serve them simultaneously at large spectral efficiencies.  


In order to achieve large spectral efficiencies in the downlink (DL) via multiuser (MU) MIMO,  channel state information at the transmitter (CSIT) is needed. Following the massive MIMO approach  \cite{marzetta-massive},  CSIT can be obtained  from the users' uplink (UL) pilots via Time-Division Duplexing (TDD) and  UL/DL radio-channel reciprocity.  This  allows training large antenna arrays by allocating as few UL pilot dimensions as the number of single-antenna users simultaneously served.  

As is well known, with isotropic channels, the number of users that can be simultaneously trained (or the system multiplexing gains) is limited by the coherence time and bandwidth of the channel \cite{caire2010multiuser},\cite{lizhong-coh}. Noting that the coherence time is  inversely proportional to the carrier frequency, increasing the carrier frequency ten-fold, e.g., from $3$ GHz to $30$ GHz, results in a ten-fold decrease in coherence time, and, thereby, in the number of user channels that can be simultaneously  trained within the coherence time of the channel.  

In this paper we focus on single-cell DL transmission over a mmWave channel, enabled by UL training and UL/DL channel reciprocity \cite{scalable-sync-cal-2014a}.   We take advantage of the sparsity of mmWave channels in the angular domain to devise schemes that yield increases in the system spatial multiplexing gains.  Indeed, typical mmWave channels are characterized by  fewer multipath components than channels at lower frequencies \cite{Bajwa-sparse}, \cite{icc2016-chenwei}, \cite{samimi},\cite{taejoon} resulting in a sparser angular support, both at the BS and the user terminal.  This channel sparsity can be exploited to train multiple user channels simultaneously, that is, training multiple users using the same pilot dimension. 
 
We consider a combination of non-orthogonal UL training from the user terminals based on pilot designs in \cite{icc2016-ozgun} and sector-based processing and precoding from the BS with the goal to increase aggregate spatial multiplexing gains and user rates. The challenge with more than one user transmitting pilots on the same pilot dimension is pilot contamination which can substantially limit massive MIMO performance, as  the beam 
used to send  data (and therefore beamforming) to one user also beamforms unintentionally at the other (contaminating) user terminal. 

In this work, multiple users, each equipped with many antenna elements and a single RF chain, are scheduled to transmit beamformed pilots on the same pilot dimension, thereby increasing the number of users simultaneously transmitting pilots for training. We exploit the presence of a massive Uniform Linear Array (ULA)  at the BS and a form of pre-sectorization in the Angle of Arrival (AoA) domain. Elementary processing at each sector allows  determining the subset of  user channels that  can be {\em resolved} on the sector, effectively pilot contamination free. Each sector then serves only the subset of users whose channels it can resolve. This allows resolving multiple users on the same pilot dimension at different sectors, thereby increasing the overall multiplexing gains of the system. 

Our approach has strong connections  but also important differences with respect to joint spatial division and multiplexing (JSDM), a two-stage method proposed in \cite{Ansuman-JTSP-2014}. JSDM partitions users into groups with approximately similar channel covariances, and exploits two-stage downlink beamforming.  In particular, precoding comprises a pre-beamformer, which depends on the user-channel covariances and minimizes interference across groups, in cascade with a MU MIMO precoder, which uses instantaneous CSI to multiplex users within a group.  Using JSDM, two users with no overlapping AoA  support in their channels can be trained and served simultaneously.   JSDM has also been studied over mmWave band channels \cite{adhikary-mmWave-jsac}; assuming full knowledge of the angular spectra of all the users, user scheduling algorithms were devised to maximize the spatial multiplexing, or received signal power. 
Our work similarly harvests  spatial multiplexing gains, but  the support of each user's spectra are not a priori known and no special scheduling is employed.

We also study how varying the user beam width can affect these harvested multiplexing gains. Indeed, using a directional beam at a user terminal makes its  user-channel sparser in terms of the number of sectors that are excited at the BS, thereby leaving more sectors available to resolve other users' channels.  Our analysis and simulations reveal that a proper choice of the user beam width can positively impact both multiplexing gains and long-term user rates.

\vspace{-1cm}
\section{System Model}
\label{sec_system_model}
\vspace{-0.45cm}
We consider a single-cell scenario, involving a single BS serving  $K_{\rm tot}$ user terminals. The BS is equipped with an $\MB$-element ULA and $\MB$ RF chains (i.e., one RF chain per antenna), while each terminal is equipped with an $\MU$-element ULA and a single RF chain. We assume OFDM and a quasistatic block fading channel model whereby the  channel of the $k$-th user stays fixed within a fading block (within the coherence time and bandwidth of the channel).  During a given fading block, the channel response between the BS and user $k$  is the $\MB \times \MU$ matrix\footnote{We assume reciprocal uplink and downlink channels hence we use $\Hm_k(f)$ for both. See \cite{scalable-sync-cal-2014a}. }   \cite{Bajwa-sparse}, \cite{sayeed-virtual}:
\begin{equation*}
\Hm_k(f) = \sum_{n = 1}^{N_p}\beta_n\avb(\thetab_n)\avu^\herm(\thetau_n)e^{-j2\pi\tau_nf},
\end{equation*}
where $N_p$ is the number of paths, and $\beta_n$ and $\tau_n$  denote the complex gain and relative delay, respectively, associated with the $n$-th path\footnote{For notational convenience, we have suppressed the dependence of $N_p$, $\beta_n$, $\tau_n$, $\thetab_{n}$ and $\thetau_{n}$ on the user index $k$.}. 
The $\MB \times 1$ vector $\avb(\theta)$ and the $\MU \times 1$ vector $\avu(\theta)$ represent the array response and steering vectors, and are $1$-periodic  in $\theta$. The normalized angle $\theta$ is related to the physical angle $\phi$ (measured with respect to array broadside) as $\theta = D\sin(\phi)$, where $D$ is the antenna spacing between two antenna elements normalized by the carrier wavelength. Assuming a maximally spread channel in angular domain, the support of both $\avb(\theta)$ and $\avu(\theta)$ are $[-1/2, 1/2]$, as in \cite{Bajwa-sparse}. 

In this paper,  we assume TDD operation and focus on  DL data transmission enabled by UL pilot transmissions from the user terminals and reciprocity-based training  \cite{marzetta-massive}.  As a result, in the case of uplink pilot (downlink data) transmission,  $\thetab_{n}$  and $\thetau_{n}$ denote the $n$-th path angles of arrival (departure) and departure (arrival).  

Spatial filtering can be applied at both the BS and the user terminal side. Given that each user terminal has a single RF chain, a user may transmit its pilot on an arbitrary  $\MU \times1$ beam $\bv$. Letting  $\alpha_{n}(\bv) = \beta_n \avu^\herm(\thetau_{n})\bv$, and using $\Pc(\bv)$ to denote the set of indices of paths that are excited with user's UL transmission via beam $\bv$, 
the physical model for the vector channel can be written as follows:
\begin{equation}
\label{hvk-def}
\hv_k(f) = \hv_k(f;\,\bv)= \sum_{n \in \Pc(\bv)}\alpha_{n}(\bv)\avb(\thetab_{n})e^{-j2\pi\tau_nf}.
\end{equation}

We let $\Rm_k \eqdef \E{\hv_k(f)\hv_k^\herm(f)}$ denote the $k$-th user channel covariance matrix and note that, due to uncorrelated scattering, $\Rm_k$ is independent of the tone index, $f$.  
Given that our focus is on the large $\MB$ case, we will assume that the DFT matrix whitens $\Rm_k$ and, as a result,  $\Rm_k$ is circulant\footnote{Indeed,  for ULAs with large $\MB$, the eigenvectors of the channel covariance matrix are accurately approximated by the columns of a DFT matrix \cite{Ansuman-JTSP-2014}.}.  Hence, the eigendecomposition of $\Rm_k$ is given by  $\Rm_k= \Fm \Lambdam_k\Fm^\herm$, with  $\Fm$ denoting the $M\times M$ DFT matrix, and $\Lambdam_k = {\rm diag}\left(\lambda_{1,k}\ldots,\lambda_{\MB,k}\right)$ where $\lambda_{1,k}\ldots,\lambda_{\MB,k}$ are the eigenvalues of $\Rm_k$.

The MU MIMO schemes we consider in this paper combine a form of spatial division and multiplexing based on instantaneous CSI.  The schemes rely on a form of pre-sectorization in the AoA domain. First
 $\hv_k(f)$ is projected onto $\Fm$ to generate the $\MB\times 1$ vector of channel observations $\gv_k(f) \eqdef\Fmb^{^\herm}\hv_k(f)$.
 Subsequently,  the $\MB$ entries of  $\gv_k(f)$ are split into $S$ non-overlaping ``sector'' groups. In particular, assuming without loss of generality, that $g=\MB/S$ is an integer, each sector comprises $g$ consecutive  entries of $\gv_k(f)$.\footnote{If $\MB$ is not divisible by $S$, groups of different sizes can be arranged.} We let $\gv_{s,k}(f)$  denote the $g\times 1$ vector associated with sector $s$ for $s  \in \{1,2,\ldots,S\}$: $\gv_{s,k}(f)$  can be expressed as 
 \begin{equation}
 \label{gvskf-def}
 \gv_{s,k}(f) = \Fmb^{^\herm}_s\hv_k(f),
 \end{equation}
  where the $\MB\times g$ matrix $\Fmb_s$ 
 comprises the $s$-th set of $g$ consecutive columns of $\Fmb$. It is worth remarking  that, since  the  entries of $ \gv_k(f) $ are uncorrelated  ($\E{\gv_k(f)\gv_k^\herm(f)} = \Lambdam_k$); in this way the $\MB \times 1$ channel vector between a single BS and a user is turned into $S$ orthogonal $g\times 1$ sector channels with uncorrelated entries. 
 
We define the average channel gain between a user $k$ and a sector $s$ as follows:
\begin{equation}\label{barlam}
\barlam{s}{k}=
 \frac{1}{g} \sum_{i = (s-1)g+1}^{ sg}\lambda_{i,k}.
\end{equation} 




In the schemes we consider, the UL pilot transmissions by the user terminals allow each sector to {\em detect} the subset of the users it sees with sufficiently high pilot SINR, and subsequently serve the associated user streams with a form of zero-forced beamforming. 
In the baseline training schemes where each user is given a dedicated UL pilot dimension and is thus not interfered by the other users' pilots, user $k$ is considered to have a high pilot SINR in  sector $s$ as long as $\barlam{s}{k}$ exceeds  some predetermined threshold $\gamma$ as the user's pilot is not interfered with other users' pilots.

We also investigate the  viability of non-orthogonal training schemes according to which  multiple users are assigned on the same UL pilot dimension. In this case, if the pilots of multiple users using the same pilot dimension are received at sufficiently high power at a given sector (i.e., UL pilots collide at this sector), none of these user channels are {\em resolvable}. 
With non-orthogonal training, user $k$ is considered resolvable in sector $s$ if no collision is declared in its dedicated pilot dimension on sector $s$,  and $\barlam{s}{k}$ exceeds $\gamma$. Details of detecting high pilot SINR (or, resolvable) users are given in Sec. \ref{training}. 

With non-orthogonal training, given $\tau$ dimensions per quasistatic fading block are used for UL training, each sector can at most serve simultaneously $\tau$ users per fading block. A user can be served by more than one sector at a time and each sector can serve more than one user at a time.\footnote{Note that, although a user's stream  transmissions  from different sectors are precoded independently, they coherently combine at the user terminal.} The instantaneous multiplexing gain over a fading block is thus the number of users that are served (by at least one sector) in that block and can {\em exceed} the available pilot dimensions, $\tau$. With orthogonal training, the multiplexing gains are upper-bounded by the number of scheduled users $\tau$.

As (\ref{hvk-def}) reveals, the choice of the beam $\bv$ employed by the user terminal to transmit its UL pilot affects $\Pc(\bv)$,  the set of indices of paths that are excited.  As explained in Appendix \ref{app-sparse}, different training beams may excite different paths but also {\em different numbers} of paths.  In fact, the sparsity of the user channel in the AoA domain (as reflected by the number of  $\bar{\lambda}_{s,k}$'s exceeding $\gamma$) can be controlled by the choice of the user {\em beam width}. In Secs.~\ref{downlink}-\ref{training} we  describe UL training and precoding schemes and tools for analyzing their performance for the case that each user employs a fixed but arbitrary beam for UL pilot transmission (and DL data reception). Subsequently, Sec. \ref{simulations} studies the effect of the beam width choice on multiplexing gains and user rates.




%

It is worth noting that \cite{adhikary-mmWave-jsac} also uses sparsity in AoA domain in the mmwave band to increase multiplexing gains. In \cite{adhikary-mmWave-jsac}, the   $\Lambdam_k$'s are assumed to be known prior to scheduling. Indeed, various user scheduling algorithms are designed to assign users to individual eigen directions based on knowledge of the $\Lambdam_k$'s.  In contrast, our work does not assume knowledge of the eigenvalues $\lambda_{i,k}$'s or $\barlam{s}{k}$'s to schedule  transmission of user streams in each sector.

\section{DL MU-MIMO Precoding}  \label{downlink}
In this section we describe the DL precoding schemes under consideration. We assume wideband scheduling, according to which a scheduling slot comprises $Q>1$ concurrent fading blocks, an let  $\tau$ denote the number of available orthogonal pilot dimensions per fading block.
Each fading block can be viewed as spanning a contiguous set of time-frequency elements in the OFDM plane that are within the coherence bandwidth and time of the user channels. Since the fading blocks  in a slot are concurrent (i.e., distinct fading blocks span distinct subbands over the same set of OFDM symbols),  we  index the fading blocks in a slot using a fading-block frequency index $f\in\{1,\,2,\cdots,\, Q\}$. With this interpretation   $\gv_{s,k}(f)$ in (\ref{gvskf-def}) denotes to the  channel of user $k$ in sector $s$ and fading block $f$.

We assume $L$  users (out  of the total of  $K_{\rm tot}$ users served by the BS) are scheduled (in round robin fashion) per slot by the BS.  In the context of the  baseline orthogonal training scheme, the BS schedules $L = \tau$ users per scheduling slot for UL pilot transmission.  Thus $\tau$ users send orthogonal pilots on each fading block, i.e., one user per pilot dimension ($K=1$). With non-orthogonal UL training, as in \cite{icc2016-ozgun}, the BS schedules $L=K\tau$ users per slot for some $K>1$.  Hence,  $K>1$ users send pilots per pilot dimension.  We use $\sigma_k$ to denote the pilot dimension used by user $k$, and $\uicode{\sigma}$ to denote the indices of users assigned to pilot dimension $\sigma$ for $1\le \sigma\le  \tau$. 

We consider DL transmission over a generic slot, and assume without loss of generality that the scheduled users have indices from $1$ to $L$. 
Assuming user $k$ uses the same beam $\bv=\bv_k$ for UL pilot transmission and as a receive front-end in the DL MIMO phase, the received signal at user $k$ over one channel use within fading block $f$ is given by 
\begin{equation}
\label{dlmimo}
r_k =\sqrt{\rho_d}  \xv^\transp(f)\,\hv_k(f) +n_k,
\end{equation}
where $\xv$ is the precoded signal, and where $n_k$ represents IID noise with $n_k\sim\Cc\Nc(0,1)$ and $\rho_d$ is the DL SNR.

\begin{figure}
\centering
\includegraphics[width=7cm, height = 3cm]{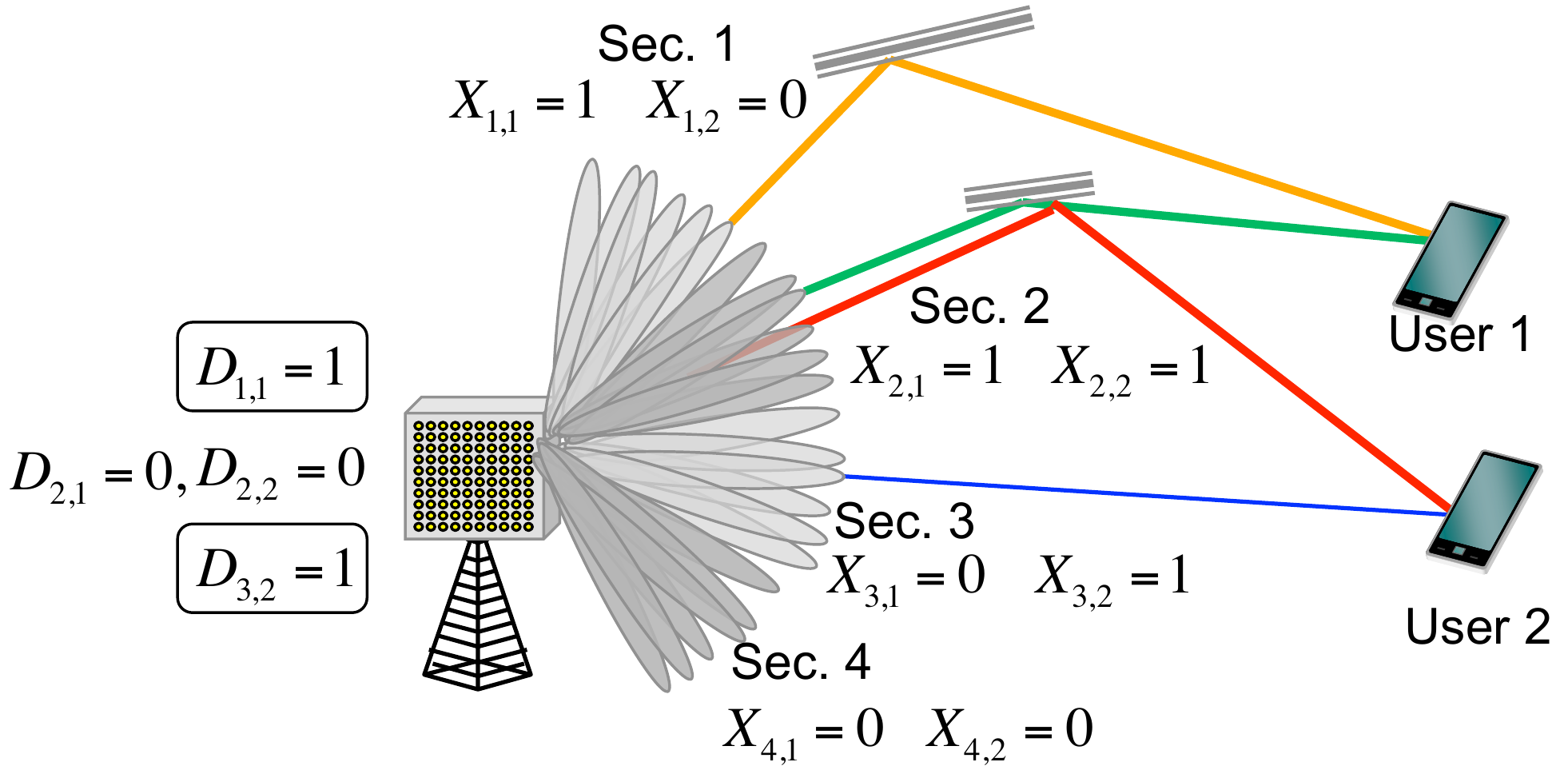}
\caption{Example of resolving two user channels on a common pilot dimension on each of the first 4 sectors of a BS.}
\label{resolvable}
\vspace{-0.4cm}
\end{figure}

In the MU-MIMO schemes we consider, precoding is sector based. In particular, based on UL training, each sector {\em resolves} the channels of a subset of the $L$ users and serves them simultaneously. We let
\begin{equation}
\label{presus-def}
\presus{s}{k} = 1_{[\gamma,\,\infty)}(\barlam{s}{k})
\end{equation}
denote whether or not user $k$ is present on sector $s$ and
\begin{equation}
\label{presec-def}
\presec{s}  = \{ k; \ \presus{s}{k} = 1\}
\end{equation}
denote the set of all users that are present in sector $s$.  In the precoding schemes we consider, user $k$ is {\em resolved} on sector $s$ (and thus will be served by sector $s$) if and only if $\presus{s}{k}=1$  and there is {\em no other} user $k'$  sharing the same dimension as user $k$ for  which $\presus{s}{k'}=1$.  Specifically we let $\detus{s}{k}$ denote whether or not user $k$'s channel can  be {\em resolved} on sector $s$:
\begin{eqnarray}
\label{detus-def}
 \detus{s}{k} &=& \presus{s}{k}\left[ \prod_{k'\in \uicode{\sigma_k}\setminus \{k\}} \left(1-\presus{s}{k'}\right) \right], \\
\label{detsec-def}
\detsec{s}  &=& \{ k; \ \detus{s}{k} = 1\}
\end{eqnarray}
be the subset of present users whose channels are resolvable in sector $s$. Fig. \ref{resolvable} shows 
an  example, involving two users using a common pilot dimension, a BS and four of its sectors, and two scatterers. As the figure reveals, user 1 is present in sectors 1 and 2, while user 2 is present in sectors 2 and 3. As a result, the channel of user 1 is resolvable in sector 1, the channel of user 2 is resolvable in sector 3, and neither user channel is resolvable in sector 2 or 4. 

In general, not all present users are resolvable and we have $\detsec{s} \subseteq \presec{s}$. Indeed, as inspection of (\ref{detus-def})  
reveals if there are two users  $k$ and $k'$  present in sector $s$  (i.e., $\presus{s}{k}  = \presus{s}{k'}=1$) that  use the same pilot dimension (i.e., with $\sigma_k  =\sigma_{k'}$), we have $\detus{s}{k}  = \detus{s}{k'}=0$. This is consistent with the fact that neither channel can be resolved  due to the pilot collision. The number of sectors that can resolve (and thus will serve) user $k$ is hence given by $N_k = \sum_{s=1}^S \detus{s}{k}$, while the number of users that are actually served in the slot is given by 
\begin{equation}
\label{MGinst-def}
L' = \left| \{ k;   N_k >0 \} \right|
\end{equation}
and, in general, $L'\le L$.

In this paper we focus on a particular form of linear zero-forced beam-forming.  All $L'$ users are given equal power, that is, power $\rho_d/L'$. Furthermore for any served user $k$, the power allocated to its stream is equally split across all sectors that resolved the user's channel. Hence, the $k$-th user's stream receives power  $\rho_d/(L'N_k) $ from each sector that serves the user. The precoded $1\times M$ signal transmitted by the BS is given by 
\begin{equation}
\label{xv-def}
\xv^\transp(f) = \sum_{s=1}^S \infosigv^\transp(f)  \Vmb^\herm_s(f)  \Fmb^\herm_s 
\end{equation}
where $\infosigv^\transp=\begin{bmatrix}  \infosig_1 &  \infosig_2 & \ldots &  \infosig_L \end{bmatrix}$ is the information bearing signal with $\infosig_k\sim\Cc\Nc(0,1)$, and 
\begin{equation}
\label{Prec-form}
\Vmb_s(f) = \begin{bmatrix}  \vvb_{s,1}(f) &  \vvb_{s,2}(f) & \ldots &  \vvb_{s,L}(f) \end{bmatrix}\end{equation}
 denotes the $g\times L$ precoder at sector $s$ and fading block $f$.  In particular,   $\vvb_{s,k}(f) = \zerov$ for any $k\notin\detsec{s}$. For any $k\in\detsec{s}$,   $\vvb_{s,k}(f)$ is in the direction of the unit-norm vector that is zero-forced to all other resolvable user-sector channel estimates, i.e., to $\{\hat{\gv}_{s,k'}(f);\ \ k' \in  \detsec{s}\setminus \{k\}\, \}$ where $\{\hat \gv_{s,k}\}$'s denote the estimates of $\{\gv_{s,k}\}$'s.  Also $\|\vvb_{s,k}(f)\|^2 = 1/(L'N_k)$. Note that with this type of precoding, $\Vmb_s(f)$ is invariant to any scalar (and complex) scalings of any of the $\hat{\gv}_{s,k}(f)$, for $k\in \detsec{s}$. Substituting the expression for $\xv(f)$ from (\ref{xv-def}) in (\ref{dlmimo}), and using the fact that $\hv_k(f) = \Fm \gv_k(f)$ we obtain
\begin{equation*}
r_k   = \sqrt{\rho_d} 
\sum_{s=1}^S \infosigv^\transp(f)  \Vmb^\herm_s(f)  \gv_{s,k} +n_k.
\end{equation*}

\section{Training, Resolvable channels and Performance Metrics}\label{training}

We next consider  orthogonal and non-orthogonal UL training  and its implications on user channel resolvability.  
Each scheduled user $k$  for $1\le k \le L$ is scheduled to transmit pilots on pilot dimension $\sigma_k$, that is, one of the $\tau$ pilot dimensions. Each pilot dimension comprises $Q$ resource elements, one per fading block.  We let $\pv_k = \begin{bmatrix} p_k(1) & p_k(2) & \cdots & p_k(Q)\end{bmatrix}^\transp$ denote the UL pilot vector transmitted by user $k$, with $p_k(f)$ denoting the pilot value used  by user $k$ on fading block $f$.

The received signal by the BS array that is based on the pilots transmitted by  the user set 
$\uicode{\sigma}$  on the pilot dimension $\sigma$ in fading block $f$    is given by
\begin{equation}
\label{yvulsig}
\yv_{\sigma}^{\rm ul}(f) = \sqrt{\rho_p} \sum_{k\in \uicode{\sigma}} \hv_k(f) p_k(f)+\wv_{\rm \sigma}^{\rm ul}(f),
\end{equation} 
where $\yv_{\rm \sigma}^{\rm ul}(f)$ is the received vector of length $\MB$, $\rho_p$ is the UL SNR, and the noise $\wv_{\sigma}^{\rm ul}$ is $\Cc\Nc(0,\Id)$. The corresponding $s$-th sector observations are given by projecting $\yv_\sigma^{\rm ul}(f)$ onto $\Fmb_{s}$
\begin{equation}
\label{yvulssig}
\bar\yv_{s,\sigma}(f) =\Fmb^\herm_{s}\yv_\sigma^{\rm ul}(f)= \sqrt{\rho_p}
\sum_{k\in \uicode{\sigma}} p_k(f) \gv_{s,k}(f) +\bar\wv_{s,\sigma}^{\rm ul},
\end{equation} 
and where $\bar\wv_{s,\sigma}^{\rm ul} =  \Fmb^\herm_s \wv_\sigma^{\rm ul}\sim\Cc\Nc(0,\Id)$, since $ \Fmb^\herm_s \Fmb_s =\Id$.

\subsection{Orthogonal training: } 

In the orthogonal training setting, user $k$ for $1\le k\le \tau$ transmits pilots on the dedicated pilot dimension  $\sigma_k=k$ (i.e., $\uicode{\sigma}=\{\sigma\}$), and, as a result, there is no user $k'\ne k$ for which $\sigma_k = \sigma_{k'}$.  Assuming  also, without loss of generality,
that  $p_k(f)=1$, the associated received signal  in fading block $f$  by the BS array based on the pilot transmitted by user $k$ on the pilot dimension $k$ (since $\sigma_k = k$) from (\ref{yvulsig}) is given by 
\begin{equation}
\label{yvulk}
\yv_k^{\rm ul}(f) = \sqrt{\rho_p} \,\hv_k(f)+\wv^{\rm ul}(f),
\end{equation} 
while the corresponding $s$-th sector observations are given by 
\begin{equation}
\label{yvulsk}
\bar\yv_{s,k}(f) = \sqrt{\rho_p}\gv_{s,k}(f) +\bar\wv_{s,k}^{\rm ul}\ .
\end{equation}


The precoder uses the following estimate of the  $k$-th user's instantaneous channel  on fading block $f$ and sector $s$:
\begin{equation}\label{orth-est2}
\hat{\gv}_{s,k}(f) = {\bar \yv}_{s,k}(f) \ .
\end{equation}
Note that this estimate does not make any use of the pilot SNR, and does not rely on knowledge of $\barlam{s}{k}$.

Inspection of (\ref{detus-def}) reveals, that  in the orthogonal scheme,   $\presus{s}{k}= \detus{s}{k} $, 
as there is no user $k'$ for which $\sigma_k = \sigma_{k'}$. That is, in the orthogonal scheme, a sector can resolve the channel of user $k$ if a user is present in sector $s$ (i.e.,  $\barlam{s,k}\ge\gamma$), and thus $\presec{s} = \detsec{s}$. 
Subsequently, the sector $s$  forms $\Vmb_s(f)$ for its resolvable user set $\detsec{s}$ according to (\ref{Prec-form}) and using $ \hat{\gv}_{s,k}(f)$ from (\ref{orth-est2})  for all $k\in \detsec{s}$.

Practical schemes for detecting the set of {\em resolvable} user channels  can be  devised by exploiting the key fact that 
\begin{equation}
\label{Eggherm-f}
\E{\!\gv_{s,k}(\!f)\gv_{s,k}^\herm(\!f)\!}\!=\! \diag\!\left(\!\lambda_{(\!s\!-\!1\!)g\!+\!1, k},  \lambda_{(\!s\!-\!1\!)g\!+\!2, k}, \ldots\!, \lambda_{sg\!, k}\!\right)\!\!
\end{equation}
for each fading block $f$ in the slot. Noting also that $\E{\|\bar \yv_{s,k}(f)\|^2}= {\rm tr}\left(\E{\bar \yv_{s,k}(f)\bar \yv_{s,k}(f)^\herm} \right)=g (\rho_p \,\,\barlam{s}{k}+1)$, we can devise simple practical detection schemes that benefit from averaging over both the $g$ beams and the $Q$ tones, e.g.:
\begin{equation}
\frac{1}{Q\rho_p g}\sum_{f=1}^Q {\|\bar \yv_{s,k}(f)\|^2} - \frac{1}{\rho_p}
\quad\mathop{\gtreqless}_{\detushat{s}{k}=0}^{\detushat{s}{k}=1}\quad \gamma.
\end{equation} If $\detushat{s}{k} = 1$, then user $k$'s channel on sector $s$ is considered as resolvable by the BS based on received UL signal.


\subsection{Non-Orthogonal training}

In the non-orthogonal training setting we consider, the pilots of $K>1$ users pilots are aligned on a single pilot dimension. As a result, the non-orthogonal scheme  splits the $L =K\tau$ scheduled users uniformly across the $\uicode{\sigma}$ sets, so that $|\uicode{\sigma}|=K$  for each $\sigma\in \{1,\, 2,\, \cdots,\, \tau\}$.
It is worth noting that, given $\presec{s}$ (the set of present users in sector $s$) the set of detected and served users from sector $s$, $\detsec{s}$, is given by (\ref{detsec-def}) and in general satisfies $\detsec{s} \subseteq \presec{s}$. For example, if there is a $\sigma$ for which multiple users are present in sector $s$, i.e.,  $|\presec{s}\cap \uicode{\sigma}|>1$ then $\detsec{s} \subset \presec{s}$. Such situation would correspond to a {\em collision}, that is, two or more users using the same pilot dimension are present in sector $s$, in which case, neither one's channel is resolvable for transmission. Given $\detsec{s}$,  sector $s$  forms $\Vmb_s(f)$ according to (\ref{Prec-form}) and using $\hat{\gv}_{s,k}(f) = \bar\yv_{s,\sigma_k}(f)$ for all $k\in \detsec{s}$ where $\bar\yv_{s,\sigma_k}(f)$ is given by (\ref{yvulssig}).

Practical detection schemes that detect which  user channels are resolvable in a sector can be also readily devised.   Noting
\begin{equation}
\label{eigen-estimate-gen}
\E{\|\bar \yv_{s,\sigma}(f)\|^2} =
g \, \left(\rho_p  \! \sum_{k\in \uicode{\sigma}} \barlam{s}{k} |p_k(f)|^2+1\right),
\end{equation}
and assuming a sufficiently large number of beams/sector, $g$, the RHS of (\ref{eigen-estimate-gen}) can be approximated by $\|\bar \yv_{s,\sigma}(f)\|^2$. This suggests that a system of $Q$ linear equations (one per fading block on pilot dimension $\sigma$) can be used to obtain $\{\barlam{s}{k}; \ \  k\in\uicode{\sigma}\}$'s.
Letting, $\secenv{s}{\sigma} =\begin{bmatrix} \| \bar \yv_{s,\sigma}(1)\|^2 & \| \bar \yv_{s,\sigma}(2)\|^2 \cdots \| \bar \yv_{s,\sigma}(Q)\|^2 
\end{bmatrix}^\transp$, we have the following system of equations: 
\begin{equation}\label{pilot-equations}
\secenv{s}{\sigma} = g \, \rho_p \, \codemat{\sigma}\,  \barlamv{s}{\sigma}  + g \onev + {\rm noise},
\end{equation} 
where $\barlamv{s}{\sigma}$ is a $K\times1$ vector whose entries comprise the set  $\{ \barlam{s}{k};\ k\in \uicode{\sigma}\}$, and where  row $f$ of the $Q\times K$ matrix $\codemat{\sigma}$ contains the associated $|p_k(f)|^2$ values. For each $k$ in $\uicode{\sigma}$, let also $i_k$ denote the index $i$ for which $[\barlamv{s}{\sigma}]_i = \barlam{s}{k}$. For $Q\ge K$, the set $\{ p_k(f); \  k\in \uicode{\sigma},\ 1\le f\le Q\}$ can be chosen a priori so that $\codemat{\sigma}$ in (\ref{pilot-equations}) has full column rank, and hence the presence of  each user can be individually detected.  One such simple detector of the presence of user $k$ is given by
\begin{equation*}
\frac{1}{\rho_p g} \ev^\transp_{i_k} \Am_{\sigma} \, \secenv{s}{\sigma} - \frac{1}{\rho_p} \ev^\transp_{i_k} \deltav_{\sigma}
\quad\mathop{\gtreqless}_{\presushat{s}{k}=0}^{\presushat{s}{k}=1}\quad \gamma,
\end{equation*} 
where $\Am_{\sigma} =  \left(\codemath{\sigma}\codemat{\sigma}\right)^{-1}\codemath{\sigma}$, and 
$\deltav_{\sigma} =  \Am_{\sigma} \onev$, and where $\ev_n$ is the $n^{\rm th}$ column of the $K\times K$ identity matrix.  Note that both $\Am_{\sigma}$ and $\deltav_{\sigma}$ are independent of the  user channels and can be computed offline. Subsequently user channel resolvability can be detected by substituting $\presushat{s}{m}$ for $\presus{s}{m}$ in (\ref{detus-def}):
\begin{equation}
\label{detush-def}
 \detushat{s}{k} = \presushat{s}{k}\left[ \prod_{k'\in \uicode{\sigma_k}\setminus \{k\}} \left(1-\presushat{s}{k'}\right) \right]\ .
\end{equation}
Various codes can be designed when $Q\ge K$ that yield $\codemat{\sigma}$ having full column rank. A full column rank matrix $\codemat{\sigma}$ allows estimating each user's large-scale response on each sector, i.e., all the $\{ \barlam{s}{k}\}$'s, thereby allowing to determine the presence of all users on all sectors.  This together with (\ref{detush-def}) allows 
detecting the users with resolvable channels.  To estimate the channels of any user $k$ that has been resolved, however, it is also necessary that $p_k(f) \neq 0$,  $\forall f\in\{1,\ldots,Q\}$. 

We remark that choosing a $\codemat{\sigma}$  that is column rank allows detecting the presence of each user but  is {\em not} necessary for detecting the resolvable user channels. Indeed, in the case that {\em multiple} active users (on a common pilot dimension $\sigma$) are present (i.e., collide) on a sector, the code design need only detect the collision event, and not the identities (and large-scale channel gains) of the  users that are present in the sector.  This fact was exploited in \cite{icc2016-ozgun} to design ON-OFF codes that are capable of resolving user channels even in cases with $Q<K$ (where $\codemat{\sigma}$  cannot be full rank).  However, pilot resource elements where a user's pilot is OFF (i.e., the $f$ values where where $p_k(f) = 0$) provide no information for channel estimation.  As a result, these ON-OFF codes incur extra pilot overheads in order to enable the BS to estimate the resolved-user channels throughout the band \cite{icc2016-ozgun}.

\subsection{Performance Metrics}

We consider two types of metrics in evaluating the performance of the proposed schemes. The first metric we use is the slot-averaged multiplexing gains provided by orthogonal and non-orthogonal training schemes: 
\begin{equation}
\label{MGinst-sample-avg}
\widehat{\rm MG} =\lim_{T\to \infty} \frac{1}{T} \sum_{t=1}^T L'(t),
\end{equation} where $L'(t)$ represents the instantaneous multiplexing gain over slot $t$ and is given by (\ref{MGinst-def}). The second performance metric is based on ergodic user-rate bounds. In Appendix \ref{Rate}, we provide closed-form rate bound expressions assuming IID channels within each sector, that is, assuming 
\begin{align}
\barlam{s}{k} = \lambda_{(s\!-\!1)g\!+\!1, k} =  \lambda_{(s\!-\!1)g\!+\!2, k} = \cdots = \lambda_{sg, k}.
\label{piecewise-lambda}
\end{align}
This abstraction is justified in Appendix \ref{Flat}.
\vspace{-0.1cm}
\subsection{Directional Training and Angular Spectra Sparsity}\label{sparse-model}
As inspection of (\ref{hvk-def}) reveals,  the number of excited paths and thus the extent to which a trained user channel is sparse in the AoA domain depends on the choice of the user beam. Similar to the AoA domain, where projecting onto the DFT basis $\Fmb$ both whitens and sparsifies the channel (that is, $\gv_k$ is both white and sparse), we consider creating the user-pilot beam as a linear combination of the AoD eigen directions,\footnote{For further details regarding how directional training sparsifies the channel in the AoA domain, see Appendix \ref{app-sparse}.} i.e., of the columns of the $\MU\times \MU$  DFT matrix,  $\Fmu$.  In particular, we consider training beams that arise from {\em activating} a subset of eigen directions.  Letting $\fvu_i$ denote the $i$-th AoD eigen direction (i.e.,  the $i$-th column of $\Fmu$), we can describe such an $\MU\times 1$ user training beam $\bv$ in terms of an $\MU\times 1$ vector $\cv$ with zero-one entries. For each $\mus$, with $1\le \mus \le \MU$,  $[\cv]_\mus$ is $1$ if $\fvu_\mus$ is activated (i.e., used as part of the training beam), while $[\cv]_\mus=0$ otherwise.  We also define the training beam width as the number of activated eigen directions, that is $\width =\onev^\transp \cv  = \sum_{\mus=1}^{\MU} \left[\cv\right]_{\mus}$. Consequently given $\cv$, the corresponding training beam $\bv$ is given by $\bv  = \bv(\cv) = \Fmu \cv/\sqrt{\width}$. Note that  $\width= 1$ corresponds to the user training on a beam with the narrowest possible width, while $\width= \MU$ corresponds to omni training.

\vspace{-0.1cm}
\section{Simulations and Conclusion} \label{simulations}
In this section,  we study the multiplexing gains and user-throughput performance of the proposed schemes with orthogonal training ($K = 1$) and non-orthogonal training ($K>1$).   In order to study the effects of  user beam width on  channel sparsity in the AoA domain and, subsequently, on multiplexing gains and user throughput, we consider a probabilistic connectivity channel model between each elemental training eigen direction in  $\{ \fvu_\mus;\ 1\le \mus \le  \MU \}$ and  each of the $S$ BS sectors. Specifically, we model the connection between BS $\text{sector}$ s and a directional beam  $\fvu_\mus$ from user $k$ as a Bernoulli random variable $\presusb{s}{k}{\mus}$ with success probability  $p$.  We also model the $\presusb{s}{k}{\mus}$'s as IID\footnote{In general, $\Pr[\presusb{s}{k}{\mus}=1] $ is, $s$, $k$, and $\mus$ dependent. In addition, $\presusb{s}{k}{\mus}$'s may be dependent random variables. Indeed, for two users $k$ and $k'$ nearby it is possible that $\presusb{s}{k}{\mus}$ and $\presusb{s}{k'}{\mus'}$ are strongly correlated for some specific training directions $\mus$ and $\mus'$.  Although important in their own right, such spatial consistency investigations are beyond the scope of this paper. For spatial consistency investigations, see \cite{asilomar2016-ozgun}. }  in $s$, $k$, and $\mus$.  

In addition, we use $\barlamb{s}{k}{\mus}$ to denote the  $\barlam{s}{k}$ induced on sector $s$ when
user $k$ training with elemental eigen direction $\bv=\fvu_{\mus}$, and model it as follows:
\begin{equation*}
\barlamb{s}{k}{\mus} \sim \begin{cases}
 {\rm U}[ \lambda_{\rm L}, \, \lambda_{\rm H}] & \text{if $\presusb{s}{k}{\mus}=1$,} \\
0 & \text{if $\presusb{s}{k}{\mus}=0$,} 
\end{cases}
\end{equation*}
for some $\lambda_{\rm L}$,  $\lambda_{\rm H}$ with $\lambda_{\rm H}>\lambda_{\rm L} >0$ and where $U[a,\, b]$ denotes a uniform distribution in $[a,\, b]$. Consequently, using a beam $\bv_k=\bv(\cv_k)$ with beam width $\width_k = \width(\cv_k)$ results in 
\begin{equation}
\label{barlamb-def}
\barlam{s}{k}= \barlam{s}{k} (\bv_k) = \frac{1}{\width_k}\sum_{\mus=1}^{\MU} \barlamb{s}{k}{\mus} [\cv_k]_{\mus}.
\end{equation}
It can be readily verified that, by choosing as a threshold  in (\ref{presus-def})  a value of $\gamma$ in the range $0< \gamma< \lambda_{\rm L}/\MU$, and using $\barlam{s}{k} (\bv_k)$  as in (\ref{barlamb-def})
the resulting $\presus{s}{k}$  in (\ref{presus-def})  satisfies $\presus{s}{k} =  \presus{s}{k}(\bv_k) 
  = \ORop_{\mus;\,  [\cv_k]_{\mus}=1} \presusb{s}{k}{\mus}$. Consequently, when user $k$ uses a given training beam $\bv_k$ of beam width $\width_k$, we have
\begin{equation}\label{prX}
\mathbb{P}\left(\presus{s}{k}=1\right) = q(\width_k),
\end{equation}
where $q(\width)\eqdef 1-(1-p)^{\width}$. Also, the $\presus{s}{k}$'s are IID in $s$, $k$.

We focus on the case where all users choose beams with the same beam width $\width$, and study the resulting multiplexing gains and user throughputs as a function of $\width$, in the range $1\le \width\le \MU$. Note that, given a common beam width $\width$,  the average number of sectors activated by a user is given by $Sq(\width)$.  Also, the expected number of scheduled users that are not present at any of the sectors is given by $L(1-q(\width))^S$.  As expected, wider beams result in broader angular support, but wider beams also make it less likely that a user is not present at any of the sectors. Using (\ref{detus-def}), (\ref{MGinst-def}), and (\ref{prX}) yields the following expression for $\widehat{\rm MG}$ in (\ref{MGinst-sample-avg}):
\begin{eqnarray}\label{MGexp1}
\widehat{\rm MG}(\width, K) &=& \E{L'}=\E{\sum_{k=1}^K\left(1- 1_{\{N_k =0 \}}\right)} \\
&=&K\, \left[1-\left(1-q(\width)\, \left[1-q(\width)\right]^{K-1}\right)^S\, \right].\nonumber
\end{eqnarray}

We next present a simulation-based study of the proposed schemes using the above model assuming  $K_{\rm tot}=100$ users terminals with $\MU = 6$ antennas each, and a BS with $M=1000$ BS antennas, using sector-based processing over $S=25$ sectors (hence $g=M/S=40$ beams per sector). We assume $p = 0.1$, and that $\tau = 5$ pilot dimensions are available for training per fading block. With these parameters, we have $q(1)=p=0.1$  and  $q(\MU) = 0.47$, implying that the  average AoA  angular support for a user ranges from 2.5 sectors  (achieved with the finest-directional training, i.e., $\width=1$) to about 11.72 sectors (achieved with omni-directional training i.e., $\width=\MU=6$). A single drop is created, i.e., a single set of  $[\presusb{s}{k}{b}]$'s are randomly created according to the model.  For any given common beam width, $\width$, at any given scheduling instance, each scheduled user picks a $\cv$ at random (out of those $\cv$'s yielding beams with beam width $\width$) . 

\begin{figure}
\centering
\includegraphics[width=7cm,height=4cm]{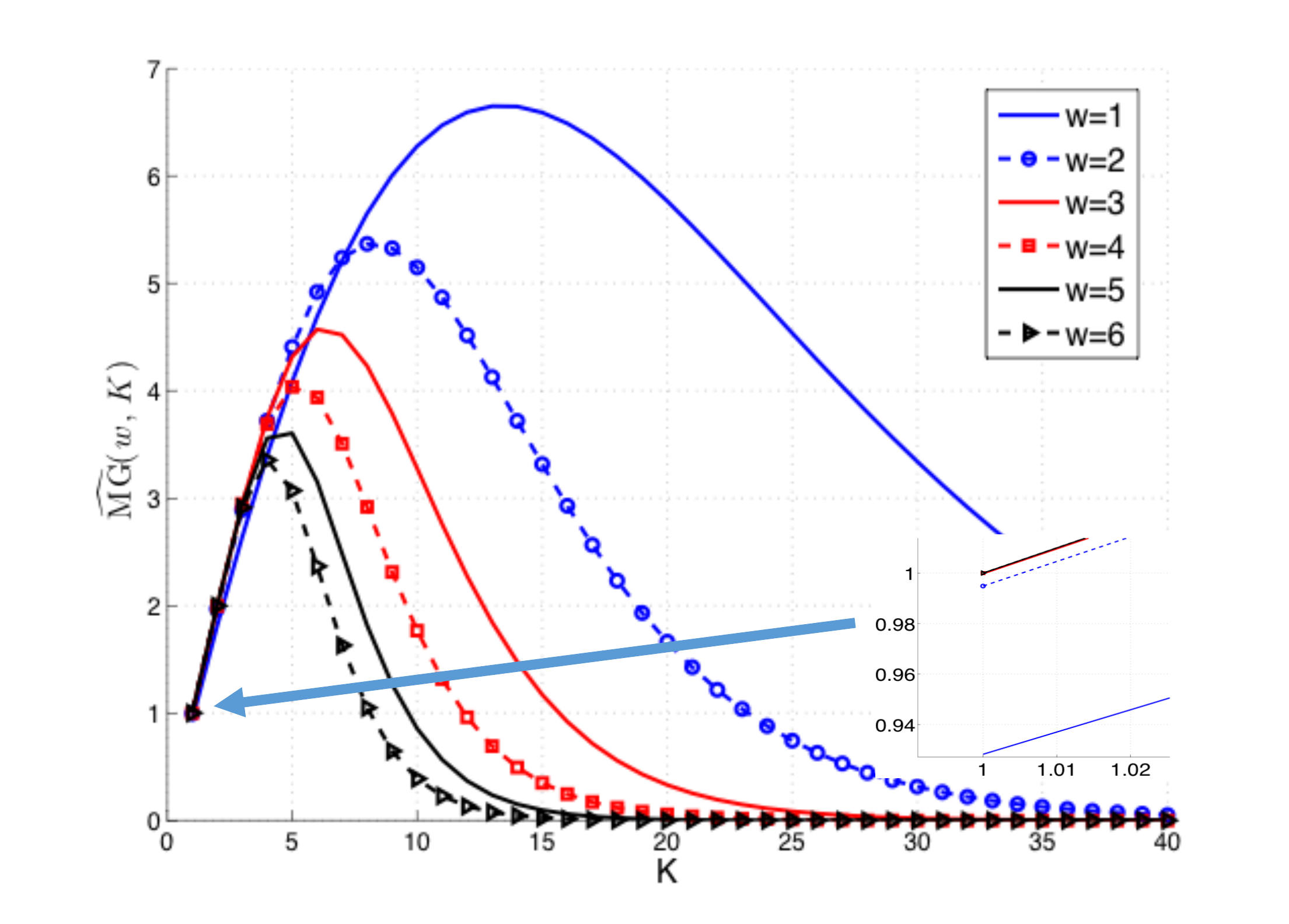}
\caption{$\widehat{\rm MG}$ vs. $K$ for different values of $\width$}
\label{Fig:MGavg_vs_K_differentw}
\vspace{-0.4cm}
\end{figure}

\begin{figure}
\centering
\includegraphics[width=9cm,height=5cm]{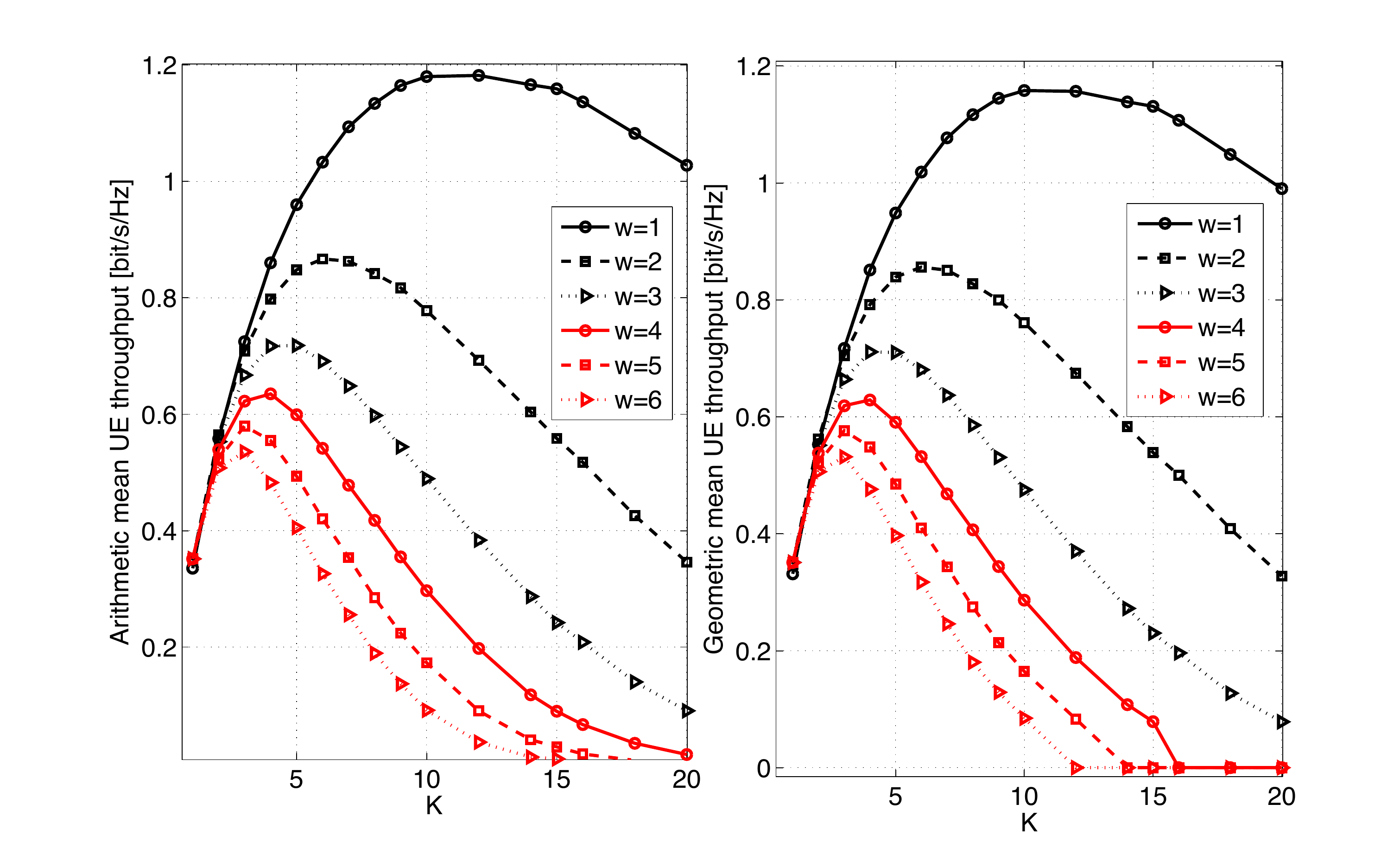}
\caption{Arithmetic/geometric mean user (UE) throughput vs. $K$ for various settings of $\width$ }
\label{Fig:AvgThroughput_vs_K}
\vspace{-0.8cm}
\end{figure}

According to (\ref{MGexp1}), Fig. \ref{Fig:MGavg_vs_K_differentw} shows the multiplexing gains per pilot dimension as a function of $K$ for different beam width values in the range $1\le \width\le\MU$. If  orthogonal training is used ($K=1$), for any beam width the  multiplexing gain is  approximately equal to (and always upper-bounded by) 1. Considering all $K\ge1$ options, for any given training-beam width $\width$, there is an optimal value of $K$, $K^{\rm opt}_{\rm MG}(\width)$, which maximizes $\widehat{\rm MG}(\width, K)$.  The best combination of beam width, $\width$, and number of scheduled users per pilot dimension, $K$,  is given by  $(\width^{\rm opt},\, K^{\rm opt})= \argmax_{(\width,K)} \widehat{\rm MG}(\width,K) = (1,\,13)$ and results in a more than 6-fold increase in multiplexing gains with respect to the orthogonal training scheme.  
With the DL SNR $\rho_d$ being $10$ dB, Fig. \ref{Fig:AvgThroughput_vs_K}a and Fig. \ref{Fig:AvgThroughput_vs_K}b, respectively, display the arithmetic and geometric mean of user throughput as a function of  $K$ for beam widths in the range $1\le \width\le\MU $.  Inspection of the two figures reveals  similar trends between the arithmetic and geometric mean of the user rates as a function of $K$ and $\width$.  Also both the arithmetic mean and the geometric mean are maximized with  $(\width,\, K)=(1,\,10)$, exhibiting in each case more than a 3-fold increase  with respect to the orthogonal training scheme. 

\begin{figure}
\centering
\includegraphics[height=2.2in]{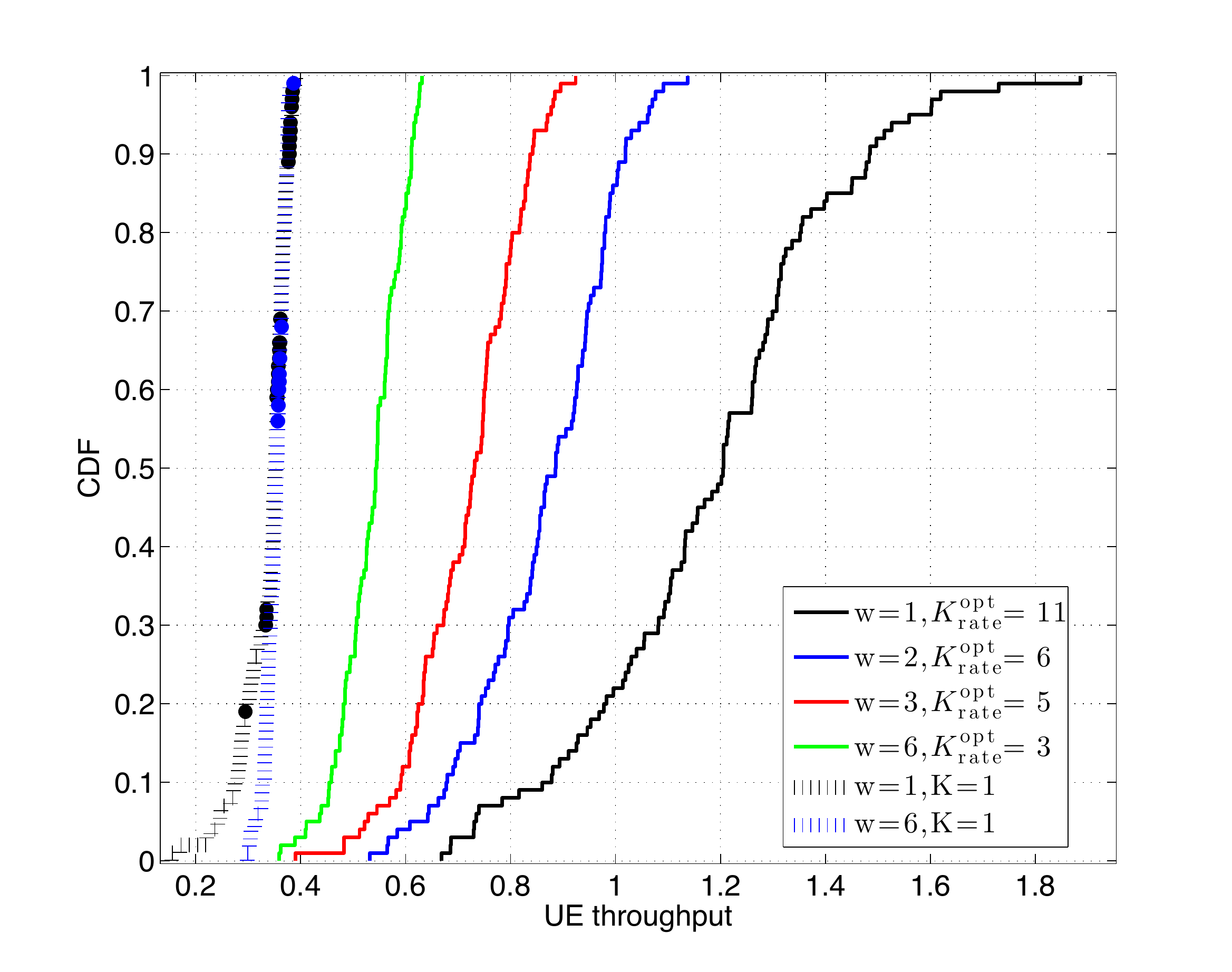}
\caption{User (UE) throughput CDF: different settings of $\width$ with corresponding $K^{\rm opt}_{\rm rate}(\width)$ }
\label{Fig:CDF_UEthroughput_differentw_Koptimal}
\vspace{-0.2cm}
\end{figure}

Fig. \ref{Fig:CDF_UEthroughput_differentw_Koptimal}  shows the empirical user-rate CDFs for all beam widths in the range $1\le \width\le \MU$. For any $\width$, the value of $K$ that 
maximizes the user-rate arithmetic mean is used. We also plot the empirical user-rate CDF with orthogonal training, for $\width=1$ and $\width=6$ (omni-training).  Inspection reveals that optimized non-orthogonal training uniformly outperforms orthogonal training in terms of individual user rates.  Furthermore, the minimum pilot beam width is best in this example, as its user-rate CDF dominates all the others. 

In conclusion, non-orthogonal UL pilots and simple large-array BS processing are jointly exploited to significantly increase cell and cell-edge throughputs over sparse mmWave channels. The proposed method leverages scheduling multiple users randomly on each available pilot dimension with random (user-chosen) training directions, and coupled with low-complexity spatial processing at the BS to resolve user channels at each BS virtual sector. Although the focus of the paper is cellular transmission and, in particular, single-cell performance, the proposed methods are directly applicable to CRAN scenarios and  cell-free type operation \cite{icc2016-ozgun,asilomar2016-ozgun}. Indeed, large improvements in multiplexing gains are also reported in the context of cell-free type networks in \cite{asilomar2016-ozgun}, based on simple, albeit spatially consistent channel models that include the effects of common scatterers and blockers. 
\vspace{-0.2cm}

\appendices
\section{Sparsity  and User Beams}\label{app-sparse}
The physical model is a very accurate representation of the real physical channel which depends on a very large number of parameters. But due to finite array apertures, limited bandwidth, $W$, the channel that is experienced in practice can be very well approximated by a discretized approximation, which is commonly known as virtual or canonical channel model \cite{Bajwa-sparse}. In fact, virtual channel model is discretized version of $\Hm_k(f)$ where uniform sampling is applied in angle-delay domains at the Nyquist rates of $1/\MB, 1/\MU , 1/W$. Virtual representation of $\Hm_k(f)$ is denoted by $\HV_k(f)$ and is given as follows \cite{Bajwa-sparse}:

\begin{eqnarray}
&&\Hm_k(f)\approx\HV_k(f) \\
&&= \sum_{i = 1}^{\MB}\sum_{m = 1}^{\MU }\sum_{\ell = 1}^{L} \Hvi_k(i,m,\ell)\avb(i/\MB)\avu^\herm(m/\MU )e^{-j2\pi\frac{\ell}{W}f},\nonumber
\end{eqnarray} where $\Hvi_k(i,m,\ell)$ is the virtual channel coefficient and it is approximately equal to the sum of complex gains of all physical paths whose angles, delays lie within the angle delay resolution bin of size $1/\MB, 1/\MU , 1/W$ and centered around the sampling point $i/\MB, m/\MU , \ell/W$, see \cite{Bajwa-sparse} for more details.

Assuming ULAs at both sides, it is well known that $\avu(m/\MU ) = \sqrt{\MU }\tilde \fv_{m}$ and $\avb(i/\MB) = \sqrt{\MB}\fv_{i}$, where $\fv_{i}$ is the $i^{\rm th}$ column of $\Fmb$ and $\tilde \fv_{m}$ is the $m^{\rm th}$ column of $\tilde \Fmb$.

\begin{figure}
\centering
\includegraphics[width=7cm, height=4cm]{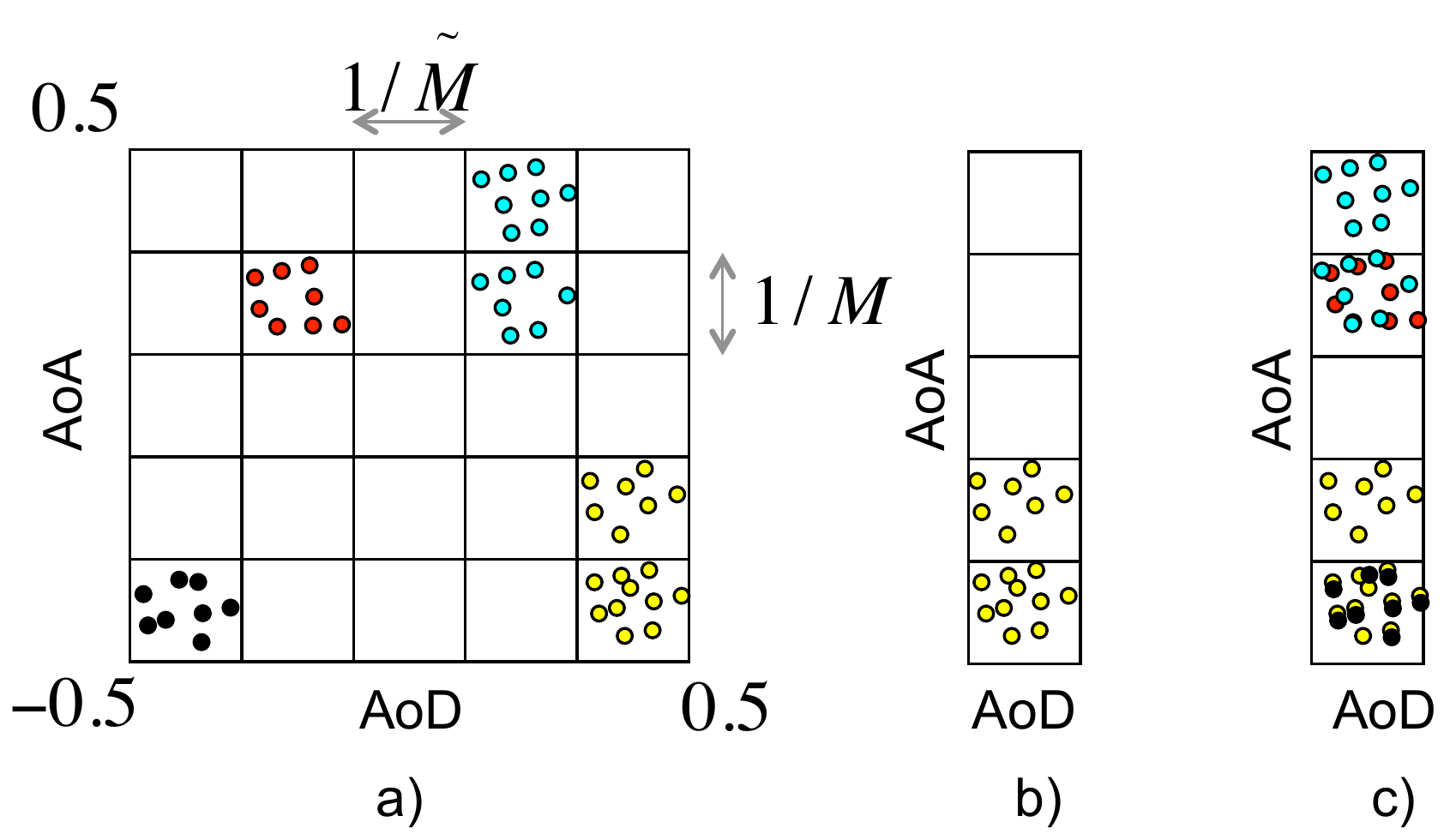}
\caption{a) An idealized sparsity pattern for AoA and AoD domain \cite{Bajwa-sparse} b) AoA sparsity with eigen beam c) AoA sparsity with omni-directional beam}
\label{fig:bajwa}
\end{figure}

To investigate the sparseness of the channel in the angle-delay domain, the virtual representation comes handy, where paths are binned together. Although the virtual channel model provides $\MB \MU W$ many bins, and hence $\MB \MU  W$ resolvable paths, in reality not necessarily all of these bins have significant paths. The sparsity of a channel can be measured by the number of bins such that $|\Hvi_k(i,m,\ell)|>\epsilon$ for some $\epsilon >0$. In Fig. \ref{fig:bajwa}, we see a sparsity pattern in angle domain. In each bin, the dots show different paths come within the specific angular bin but possibly with different delays. The bins with no paths represent the bins who virtual channels absolute value is below the threshold, ie. $|\Hvi_k(i,m,\ell)|<\epsilon$.

Next, we show how the effective sparsity of the channel is effected by the beam selection at the user side. Let $\bv$ be the user beam of size $\MU \times 1$ and unit norm, then the $\MB\times \MU $ physical channel $\HV_k(f)$ effectively becomes an $\MB\times 1$ vector channel $\hV_k(f;\bv)\eqdef\HV_k(f)\bv$.
Assuming UL, we investigate the dependency of sparsity in AoA domain to the user beam $\bv$: Consider two different extreme cases:
\begin{itemize}
\item Eigen beam: $\bv = \tilde\fv_{m}$\\
$\hV_k(f;\bv) = \sum_{i = 1}^{\MB}\avb(\frac{i}{\MB})\sum_{\ell = 1}^{L}  \Hvi_k(i,m,\ell)e^{-j2\pi\frac{\ell}{W}f}, $ \\
\item Omni-directional beam: $\bv = \frac{1}{\sqrt{\MU }}\sum_{m = 1} ^{\MU }\tilde \fv_{m}$ \\
$\hV_k(f;\bv) =\sum_{i = 1}^{\MB}\avb(\frac{i}{\MB})\sum_{m= 1}^{\MU }\sum_{\ell = 1}^{L}  \Hvi_k(i,m,\ell)e^{-j2\pi\frac{\ell}{W}f}. $ 
\end{itemize}

It is possible to see that in case of an eigen beam, in the AoA domain, bins are filled up with paths that fall into the $m^{\rm th}$ bin in AoD. On the other hand, in case of omni-directional beam, AoA bins are filled up with paths that are coming from any angle in AoD. See Fig. \ref{fig:bajwa} b) and c) for the sparsity comparison of these two cases. Vector channel is more sparse in AoA domain when eigen beam is used, compared to that of omni-directional beam. One can also imagine other beam options which can be written as a weighted sum of DFT columns. Hence by using various beams for users, we can effectively, create different random sparsity columns in AoA.

\section{Rate Derivation}\label{Rate}
With the spatial filtering, we treat each sector as an individual virtual BS. Therefore, in this part we provide a general rate bound derivation for the multi-cell distributed MIMO. The resulting expressions provide the required ergodic-rate bounds  for the single-cell sector-based schemes we consider in this paper, as a special case.

Consider $S$ BSs (sectors), where each has $g$ antennas, serving $K$ single-antenna UEs,\footnote{In our proposed system, after the phase of prebeamforming\cite{adhikary-mmWave-jsac} channel dimension is reduced to the number of RF chains. Thus, the derivation in this part fits the scenario where UE is equipped with many antennas but a single RF chain.} we assume that BSs only share the transmitted data and form beamformers individually, which alleviates the burden of instantaneous CSI exchange. 

Let's first investigate the phase of uplink training and channel estimation. Consider a general random pilot reuse scheme, where $K$ UEs are randomly partitioned into $\tau$ groups and UEs in the same group share the common pilot sequence. The received $g\times \tau$ training matrix at the $\text{BS}_{s}$ is
\begin{equation}
\mathbf{Y}_s^{ul}=\sqrt{\tau\rho_p}\sum\limits_{k=1}^K \mathbf{g}_{s,k}\boldsymbol{\varphi}_{G(k)}^\herm+\bar{\mathbf{W}}^{ul}_s,\nonumber
\end{equation}
where $\gv_{s,k}\sim\bar\lambda_{s,k}\Cc(0,\Id_{g\times g})$ is the $g$-by-$1$ transfer channel between $\text{BS}_s$ and $\text{UE}_k$, $G(k)\in[1,2,...,\tau]$ denotes the group index of $\text{UE}_k$. $[\boldsymbol{\varphi}_i]_{i=1}^\tau$ are orthogonal pilot codes and $\|\boldsymbol{\varphi}_i\|=1, \forall i$. Entries of $\bar{\mathbf{W}}^{ul}_{s}\in\mathbb{C}^{g\times\tau}$ are IID zero-mean unit-variance complex Gaussian variables. Multiplicative scalars $\tau$ and $\rho_p$ indicate processing gain and uplink SNR, respectively. 

$\text{BS}_s$ projects its received training matrix on different pilot codes and obtains observations from different pilot codes as
\begin{equation}
\bar{\mathbf{y}}_{l,s}^{ul}\defines\mathbf{Y}_s^{ul}\boldsymbol{\varphi}_l=\sqrt{\tau\rho_p}\sum\limits_{G(k)=l}\mathbf{g}_{s,k}+\bar{\mathbf{w}}_{l,s}^{ul}, \forall l,\nonumber
\end{equation}
where $\bar{\mathbf{w}}_{l,s}^{ul}=\bar{\mathbf{W}}_s^{ul}\boldsymbol{\varphi}_l\sim\Cc(0,\Id_{g\times g})$ is the effective noise vector. With the IID channel assumption, we obtain the minimum mean square error (MMSE) channel estimation, its second moment and the variance of MMSE estimation error, respectively:  
\begin{align}
&\hat{\mathbf{g}}_{s,k}\!\!=\!\!\frac{\sqrt{\tau\rho_p}\bar\lambda_{s,k}}{\tau\rho_p\sum_{G(u)\!=\!G(k)}\bar\lambda_{s,u}+1}\bar{\mathbf{y}}_{G(k),s}^{ul}\!\!=\!\!\frac{\gamma_{s,k}}{\sqrt{\tau\rho_p}\bar\lambda_{s,k}}\bar{\mathbf{y}}_{G(k),s}^{ul},\label{Eq.channelEst}\\
&\gamma_{s,k}\!\!\defines\!\!\frac{\mathbb{E}[|\hat{\mathbf{g}}_{s,k}|^2]}{g}\!\!=\!\!\frac{\tau\rho_p\bar\lambda_{s,k}^2}{\tau\rho_p\sum_{G(u)=G(k)}\bar\lambda_{s,u}+1},\nonumber\\
&\delta^2_{e,k,s}\eqdef\frac{\mathbb{E}[\|\hat{\mathbf{e}}_{s,k}\|^2]}{g}=\barlam{s}{k}-\gamma_{s,k},\nonumber
\end{align}
where $\hat{\mathbf{e}}_{s,k}\sim\delta^2_{e,k,s}\Cc(0,\Id_{g\times g})$ is the estimation error vector. According to the principle of orthogonality, $\mathbf{g}_{s,k}$ can be expressed as $\mathbf{g}_{s,k}=\hat{\mathbf{g}}_{s,k}+\hat{\mathbf{e}}_{s,k}$, while $\hat{\mathbf{g}}_{s,k}$ and $\hat{\mathbf{e}}_{s,k}$ are independent of each other. 

At the downlink transmission phase, the received signal $r_k$ at $\text{UE}_k$ is
\begin{align}
r_k&\!\!=\!\!\sqrt{\rho_d}\sum\limits_{s=1}^S \eta_{s,k}^{\frac{1}{2}}\mathbf{g}_{s,k}^\herm\mathbf{v}_{s,k}x_k\!\!+\!\!\sqrt{\rho_d}\sum\limits_{s=1}^S\sum\limits_{u\neq k}\eta_{s,u}^{\frac{1}{2}} \mathbf{g}_{s,k}^\herm\mathbf{v}_{s,u}x_u\!\!+\!\!n_k.
\label{Eq.recSig_1}
\end{align}
Since the received signal (\ref{Eq.recSig_1}) is a combination of desired signal and interference plus noise, we can represent it as
\begin{align}
r_k&=D_kx_k+z_k,
\label{Eq.recSig_2}
\end{align}
where
\begin{align}
D_k&=\sqrt{\rho_d}\sum\limits_{s=1}^S \eta_{s,k}^{\frac{1}{2}}\mathbf{g}_{s,k}^\herm\mathbf{v}_{s,k},\nonumber\\
z_k&=\sqrt{\rho_d}\sum\limits_{u\neq k}I_{k,u}x_u+n_k,\nonumber\\
I_{k,u}&=\sum_{s=1}^S\eta_{s,u}^{\frac{1}{2}} \mathbf{g}_{s,k}^\herm\mathbf{v}_{s,u}.\nonumber
\end{align} 
Based on (\ref{Eq.recSig_2}), we develop rate bound approximations to evaluate the system performance.

By assuming UE only has the knowledge of channel statistics for decoding \cite{hassibi-hochwald}, \cite{Interdonato2016} utilizes (\ref{Eq.rateBound}) for rate evaluation:
\begin{align}
R_k&=\log{(1+\text{SINR})}=\log{\big(1+\frac{|\mathbb{E}[D_k]|^2}{\var{(D_k)}+\delta_{z_k}^2}\big)},
\label{Eq.rateBound}
\end{align} 
where $\mathbb{E}[D_k]$ denotes the mean of desired signal, $\var(D_k)$ is the variance of $D_k$, indicating the self-interference brought by unknowing of instantaneous CSI, and $\delta_{z_k}^2$ is the variance of inter-user interference plus noise. However, in a practical system with finite array size, the self-interference, dominating at high SNR regime, will make ergodic rate approximation by (\ref{Eq.rateBound}) too conservative. Therefore, in this paper, we focus on the following rate approximation \cite{Shamai2002}:
\begin{align}
\bar{R}_k=\log(1+\frac{\mathbb{E}[|D_k|^2]}{\delta_{z_k}^2})-\frac{1}{T_d}\log(1+\frac{\mathbb{E}[|D_k|^2]T_d}{\delta_{z_k}^2}),\label{Eq.rateBound_caire}
\end{align}
where $T_d$ is the coherence time. Later, we will develop expressions for the terms, including $\mathbb{E}[D_k],\mathbb{E}[|D_k|^2]$ and $\delta_{z_k}^2$,which can be also used to compute the bound (\ref{Eq.rateBound}). 

In this paper, we consider the normalized zero-forcing beamforming (ZFBF) for downlink transmission, where $\|\mathbf{v}_{s,u}\|=1$, $\forall s, u$.\footnote{\cite{Interdonato2016} exhibits the derivations for the rate bound (\ref{Eq.rateBound}) with the normalized conjugate beamforming (CBF).} An interesting observation of (\ref{Eq.channelEst}) is that estimated channels of different UEs from the same group align with the observation signal on the signal space. Therefore, $\spanZ([\hat{\mathbf{g}}_{s,k}]_{k=1}^K)=\spanZ([\bar{\mathbf{y}}_{l,s}^{ul}]_{l=1}^\tau), \forall s$, and we can directly perform ZFBF on the space spanned by training observations from different pilot codes.

Without loss of generality, let's stick with the beamforming vector $\mathbf{v}_{s,k}$ formed by $\text{BS}_s$ to serve $\text{UE}_k$. Assume that $\text{BS}_s$ uses observations from a subset of pilot codes, i.e., $\mathcal{K}_s$ whose cardinality is $K_s=|\mathcal{K}_s|\leq \tau$, to form zero-forcing precoder. Note that $\mathcal{K}_s$ contains pilot codes not only corresponding to served UEs of $\text{BS}_s$ but also those  UEs, to whom $\text{BS}_s$ will null out leakage interference. 

Define matrix of interference channel 
\begin{align}
\mathbf{B}_{s,k}\eqdef[\bar{\mathbf{y}}_{l_1,s}^{ul},...,\bar{\mathbf{y}}_{l_{K_s},s}^{ul}]\setminus\bar{\mathbf{y}}_{l,s}^{ul},\nonumber
\end{align}
where $\mathcal{K}_s=[l_{i}]_{i=1}^{K_s}$ and $l=G(k)\in\mathcal{K}_s$. The beamforming vector $\mathbf{v}_{s,k}$ becomes
\begin{align}
\mathbf{v}_{s,k}&=\frac{\mathbf{\Pi}(\mathbf{B}_{s,k})\hat{\mathbf{g}}_{s,k}}{\|\mathbf{\Pi}(\mathbf{B}_{s,k})\hat{\mathbf{g}}_{s,k}\|},\nonumber\\
\mathbf{\Pi}(\mathbf{B}_{s,k})&\eqdef\mathbf{I}_{g\times g}-\mathbf{B}_{s,k}(\mathbf{B}_{s,k}^\herm\mathbf{B}_{s,k})^{-1}\mathbf{B}_{s,k}^\herm,\nonumber
\end{align}
where $\mathbf{\Pi}(\mathbf{B}_{s,k})$ is a projection matrix denoting the null space of $\mathbf{B}_{s,k}$. The mean of desired signal becomes
\begin{align}
\mathbb{E}[D_k]&=\sqrt{\rho_d}\sum\limits_{s=1}^S\eta_{s,k}^{\frac{1}{2}}\mathbb{E}[(\hat{\mathbf{g}}_{s,k}^\herm+\hat{\mathbf{e}}_{s,k}^\herm)\frac{\mathbf{\Pi}(\mathbf{B}_{s,k})\hat{\mathbf{g}}_{s,k}}{\|\mathbf{\Pi}(\mathbf{B}_{s,k})\hat{\mathbf{g}}_{s,k}\|}]\nonumber\\
&=\sqrt{\rho_d}\sum\limits_{s=1}^S\eta_{s,k}^{\frac{1}{2}}\mathbb{E}[\mathbb{E}[\frac{\hat{\mathbf{g}}_{s,k}^\herm\mathbf{\Pi}(\mathbf{B}_{s,k})\hat{\mathbf{g}}_{s,k}}{\|\mathbf{\Pi}(\mathbf{B}_{s,k})\hat{\mathbf{g}}_{s,k}\|}|\mathbf{B}_{s,k}]]\label{Eq.meanS_1}\\
&=\sqrt{\rho_d}\sum\limits_{s=1}^S\eta_{s,k}^{\frac{1}{2}}\mathbb{E}[\mathbb{E}[\|\mathbf{\Pi}(\mathbf{B}_{s,k})\hat{\mathbf{g}}_{s,k}\||\mathbf{B}_{s,k}]].\label{Eq.meanS_2}
\end{align} 
In (\ref{Eq.meanS_1}), since $\hat{\mathbf{e}}_{s,k}$ is independent of $[\hat{\mathbf{g}}_{s,k}]_{k=1}^K$, we strike out the term related to it and apply the chain rule of expectation. Substitute (\ref{Eq.meanS2_1}) into (\ref{Eq.meanS_1}), we obtain (\ref{Eq.meanS_2}).  
\begin{align}
\hat{\mathbf{g}}_{s,k}^\herm\mathbf{\Pi}(\mathbf{B}_{s,k})\hat{\mathbf{g}}_{s,k}&=\hat{\mathbf{g}}_{s,k}^\herm\mathbf{\Pi}(\mathbf{B}_{s,k})^\herm\mathbf{\Pi}(\mathbf{B}_{s,k})\hat{\mathbf{g}}_{s,k}\nonumber\\
&=\|\mathbf{\Pi}(\mathbf{B}_{s,k})\hat{\mathbf{g}}_{s,k}\|^2.
\label{Eq.meanS2_1}
\end{align}  
The singular value decomposition of $\mathbf{B}_{s,k}$ is $\mathbf{B}_{s,k} = \mathbf{U}_{s,k}\mathbf{\Lambda}_{s,k}\mathbf{V}_{s,k}^\herm$, where $\mathbf{U}_{s,k}\in\mathbb{C}^{g\times (K_s-1)}$ is semi-unitary.  We can represent $\mathbf{\Pi}(\mathbf{B}_{s,k})$ as 
\begin{align}
\mathbf{\Pi}(\mathbf{B}_{s,k})&=\mathbf{I}_{g\times g}-\mathbf{U}_{s,k}\mathbf{U}_{s,k}^\herm=\bar{\mathbf{U}}_{s,k}\bar{\mathbf{U}}_{s,k}^\herm,
\label{Eq.nullSpaceII}
\end{align}
where semi-unitary matrix $\bar{\mathbf{U}}_{s,k}\in\mathbb{C}^{g\times(g-K_s+1)}$ indicates the complement space of $\mathbf{U}_{s,k}$. Substituting (\ref{Eq.nullSpaceII}) into (\ref{Eq.meanS_2}), we first evaluate the inner conditional expectation
\begin{align}
&\mathbb{E}[\|\mathbf{\Pi}(\mathbf{B}_{s,k})\hat{\mathbf{g}}_{s,k}||\big|\mathbf{B}_{s,k}]\nonumber\\
=&\mathbb{E}[\sqrt{\hat{\mathbf{g}}_{s,k}^\herm\mathbf{\Pi}(\mathbf{B}_{s,k})\hat{\mathbf{g}}_{s,k}}|\mathbf{B}_{s,k}]\nonumber\\
=&\mathbb{E}[\sqrt{\hat{\mathbf{g}}_{s,k}^\herm\bar{\mathbf{U}}_{s,k}\bar{\mathbf{U}}_{s,k}^\herm\hat{\mathbf{g}}_{s,k}}|\mathbf{B}_{s,k}]\nonumber\\
=&\mathbb{E}[\sqrt{\tilde{\mathbf{g}}_{s,k}^\herm\tilde{\mathbf{g}}_{s,k}}|\mathbf{B}_{s,k}]\nonumber\\
=&\mathbb{E}[\|\tilde{\mathbf{g}}_{s,k}\||\mathbf{B}_{s,k}],\nonumber
\end{align}   
where we define $\tilde{\mathbf{g}}_{s,k}\defines\bar{\mathbf{U}}_{s,k}^\herm\hat{\mathbf{g}}_{s,k}$. Since $\hat{\mathbf{g}}_{s,k}$ is independent of $\mathbf{B}_{s,k}$, $\tilde{\mathbf{g}}_{s,k}$ follows $\gamma_{s,k}\Cc(0,\Id_{(g-K_s+1)\times(g-K_s+1)})$  for a given $\mathbf{B}_{s,k}$. Therefore, the conditional expectation becomes the mean of a random variable with chi-distribution:
\begin{align}
\mathbb{E}[\|\mathbf{\Pi}(\mathbf{B}_{s,k})\hat{\mathbf{g}}_{s,k}\||\mathbf{B}_{s,k}]=\sqrt{\gamma_{s,k}}\frac{\Gamma(g-K_s+\frac{3}{2})}{\Gamma(g-K_s+1)},
\label{Eq.condExp}
\end{align}
where $\Gamma(\cdot)$ is the Gamma function defined as $\Gamma(t)=\int_{0}^\infty x^{t-1}e^{-x}dx$. From (\ref{Eq.condExp}), we can observe that the conditional expectation is independent of realizations of $\mathbf{B}_{s,k}$. Therefore, we obtain the mean of desired signal
\begin{align}
\mathbb{E}[D_k]&=\sqrt{\rho_d}\sum\limits_{s=1}^S\eta_{s,k}^\frac{1}{2}\gamma_{s,k}^{\frac{1}{2}}\frac{\Gamma(g-K_s+\frac{3}{2})}{\Gamma(g-K_s+1)}\nonumber\\
&=\sqrt{\rho_d}\sum_{s=1}^S\eta_{s,k}^{\frac{1}{2}}\gamma_{s,k}^{\frac{1}{2}}\Omega(g,K_s),\nonumber
\end{align}
where we define function $\Omega(g,K_s)\defines \frac{\Gamma(g-K_s+\frac{3}{2})}{\Gamma(g-K_s+1)}$.

For the second moment of desired signal, we have
\begin{align}
&\mathbb{E}[|D_k|^2]=\rho_d\mathbb{E}[|\sum\limits_{s=1}^S\eta_{s,k}^{\frac{1}{2}}(\hat{\mathbf{g}}_{s,k}^\herm+\hat{\mathbf{e}}_{s,k}^\herm)\frac{\mathbf{\Pi}(\mathbf{B}_{s,k})\hat{\mathbf{g}}_{s,k}}{\|\mathbf{\Pi}(\mathbf{B}_{s,k})\hat{\mathbf{g}}_{s,k}\|}|^2]\nonumber\\
&=\rho_d\mathbb{E}[|\sum\limits_{s=1}^S\eta_{s,k}^{\frac{1}{2}}(\|\mathbf{\Pi}(\mathbf{B}_{s,k})\hat{\mathbf{g}}_{s,k}\|+\frac{\hat{\mathbf{e}}_{s,k}^\herm\mathbf{\Pi}(\mathbf{B}_{s,k})\hat{\mathbf{g}}_{s,k}}{\|\mathbf{\Pi}(\mathbf{B}_{s,k})\hat{\mathbf{g}}_{s,k}\|})|^2]\nonumber\\
&=\rho_d\mathbb{E}[|\sum\limits_{s=1}^{S}\eta_{s,k}^{\frac{1}{2}}(\|\bar{\mathbf{g}}_{s,k}\|+\frac{\hat{\mathbf{e}}_{s,k}^\herm\bar{\mathbf{g}}_{s,k}}{\|\bar{\mathbf{g}}_{s,k}\|})|^2]\label{Eq.2ndMoment_0}\\
&=\rho_d\sum\limits_{j=1}^S\sum\limits_{i=1}^S \eta_{j,k}^{\frac{1}{2}}\eta_{i,k}^{\frac{1}{2}}\mathbb{E}[\|\bar{\mathbf{g}}_{j,k}\|\|\bar{\mathbf{g}}_{i,k}\|+\frac{\hat{\mathbf{e}}_{j,k}^\herm \bar{\mathbf{g}}_{j,k} \bar{\mathbf{g}}_{i,k}^\herm\hat{\mathbf{e}}_{i,k}}{\|\bar{\mathbf{g}}_{j,k}\|\|\bar{\mathbf{g}}_{i,k}\|}],
\label{Eq.2ndMoment_1}
\end{align}
where we define the unnormalized beamforming vector $\bar{\mathbf{g}}_{s,k}\defines\mathbf{\Pi}(\mathbf{B}_{s,k})\hat{\mathbf{g}}_{s,k}$ to obtain (\ref{Eq.2ndMoment_0}). From (\ref{Eq.2ndMoment_0}) to (\ref{Eq.2ndMoment_1}), we strike out cross terms with mean zero. Then, consider terms where $i=j$ in (\ref{Eq.2ndMoment_1}), we have 
\begin{align}
\rho_d\sum\limits_{s=1}^S\eta_{s,k}\{\mathbb{E}[\|\bar{\mathbf{g}}_{s,k}\|^2]+\mathbb{E}[\frac{\bar{\mathbf{g}}_{s,k}^\herm}{\|\bar{\mathbf{g}}_{s,k}\|}\hat{\mathbf{e}}_{s,k}\hat{\mathbf{e}}_{s,k}^\herm\frac{\bar{\mathbf{g}}_{s,k}}{\|\bar{\mathbf{g}}_{s,k}\|}]\}.
\label{Eq.2ndMoment_4}
\end{align} 
Given $\!\mathbf{B}_{s,k}$, $\bar{\mathbf{g}}_{s,k}\!\!\!=\!\!\!\bar{\mathbf{U}}_{s,k}\bar{\mathbf{U}}_{s,k}^\herm\hat{\mathbf{g}}_{s,k}$ following $\gamma_{s,k}\Cc(0,\bar{\mathbf{U}}_{s,k}\bar{\mathbf{U}}_{s,k}^\herm)$. Therefore, we have
\begin{align}
\mathbb{E}[\|\bar{\mathbf{g}}_{s,k}\|^2]&=\mathbb{E}[\mathbb{E}[\|\bar{\mathbf{g}}_{s,k}\|^2|\mathbf{B}_{s,k}]]=(g-K_s+1)\gamma_{s,k}. 
\label{Eq.2ndMoment_5}
\end{align}

To calculate the second term in (\ref{Eq.2ndMoment_4}), we substitute $\mathbf{v}_{s,k}=\frac{\bar{\mathbf{g}}_{s,k}}{\|\bar{\mathbf{g}}_{s,k}\|}$ and get
\begin{align}
&\mathbb{E}[\frac{\bar{\mathbf{g}}_{s,k}^\herm}{\|\bar{\mathbf{g}}_{s,k}\|}\hat{\mathbf{e}}_{s,k}\hat{\mathbf{e}}_{s,k}^\herm\frac{\bar{\mathbf{g}}_{s,k}}{\|\bar{\mathbf{g}}_{j,k}\|}]=\sum\limits_{q=1}^g\mathbb{E}[|\mathbf{v}_{s,k}(q)|^2\cdot |\hat{\mathbf{e}}_{s,k}(q)|^2]\nonumber\\
=&\sum\limits_{q=1}^g\mathbb{E}[|\mathbf{v}_{s,k}(q)|^2]\mathbb{E}[|\hat{\mathbf{e}}_{s,k}(q)|^2]=\sigma_{e,k,s}^2,
\label{Eq.2ndMoment_6}
\end{align} 
where $\mathbf{v}(q)$ indicates the $q$-th element of a vector $\mathbf{v}$. 
Substituting (\ref{Eq.2ndMoment_5}) and (\ref{Eq.2ndMoment_6}) into (\ref{Eq.2ndMoment_4}), we obtain
\begin{align}
\rho_d\sum\limits_{s=1}^S\eta_{s,k}((g-K_s+1)\gamma_{s,k}+\sigma_{e,k,s}^2).
\label{Eq.2ndMoment_7}
\end{align} 
Consider terms where $i\neq j$ in (\ref{Eq.2ndMoment_1}), we have
\begin{align}
&\rho_d\sum\limits_{j=1}^S\sum\limits_{i\neq j}^S\eta_{j,k}^{\frac{1}{2}}\eta_{i,k}^{\frac{1}{2}}\mathbb{E}[\|\bar{\mathbf{g}}_{j,k}\|\|\bar{\mathbf{g}}_{i,k}||+\frac{\hat{\mathbf{e}}_{j,k}^\herm \bar{\mathbf{g}}_{j,k} \bar{\mathbf{g}}_{i,k}^\herm\hat{\mathbf{e}}_{i,k}}{\|\bar{\mathbf{g}}_{j,k}\|\|\bar{\mathbf{g}}_{i,k}\|}]\nonumber\nonumber\\
=&\rho_d\sum\limits_{j=1}^S\sum\limits_{i\neq j}^S \eta_{j,k}^{\frac{1}{2}}\eta_{i,k}^{\frac{1}{2}}\mathbb{E}[\|\bar{\mathbf{g}}_{j,k}\|]\mathbb{E}[\|\bar{\mathbf{g}}_{i,k}\|]\nonumber\\
=&\rho_d\sum\limits_{j=1}^S\sum\limits_{i\neq j}^S \eta_{j,k}^{\frac{1}{2}}\eta_{i,k}^{\frac{1}{2}}\gamma_{j,k}^\frac{1}{2}\gamma_{i,k}^\frac{1}{2}\frac{\Gamma(g-K_j+\frac{3}{2})}{\Gamma(g-K_j+1)}\frac{\Gamma(g-K_i+\frac{3}{2})}{\Gamma(g-K_i+1)}\nonumber\\
=&\rho_d\sum\limits_{j=1}^S\sum\limits_{i\neq j}^S \eta_{j,k}^{\frac{1}{2}}\eta_{i,k}^{\frac{1}{2}}\gamma_{j,k}^\frac{1}{2}\gamma_{i,k}^\frac{1}{2}\Omega(g,K_j)\Omega(g,K_i).
\label{Eq.2ndMoment_8}
\end{align}
Combining (\ref{Eq.2ndMoment_7}) and (\ref{Eq.2ndMoment_8}), we develop the expression for the second moment of the useful signal
\begin{align}
\mathbb{E}[|D_k|^2]&=\rho_d[\sum\limits_{s=1}^S\eta_{s,k}((g-K_s+1)\gamma_{s,k}+\sigma_{e,k,s}^2)+\nonumber\\
&\sum\limits_{j=1}^S\sum\limits_{i\neq j}^S\eta_{j,k}^{\frac{1}{2}}\eta_{i,k}^{\frac{1}{2}}\gamma_{j,k}^\frac{1}{2}\gamma_{i,k}^\frac{1}{2}\Omega(g,K_j)\Omega(g,K_i)].
\label{Eq.userfulSinal2nd}
\end{align}
Now let's consider interference terms. We need to find
\begin{align}
\sigma_{z_k}^2&=\rho_d\sum\limits_{u\neq k}\mathbb{E}[|I_{k,u}|^2]+\delta^2,\nonumber\\
\mathbb{E}[|I_{k,u}|^2]&=\mathbb{E}[|\sum\limits_{s=1}^S\eta_{s,u}^{\frac{1}{2}}(\hat{\mathbf{g}}_{s,k}^\herm+\hat{\mathbf{e}}_{s,k}^\herm)\frac{\mathbf{\Pi}(\mathbf{B}_{s,u})\hat{\mathbf{g}}_{s,u}}{\|\mathbf{\Pi}(\mathbf{B}_{s,u})\hat{\mathbf{g}}_{s,u}\|}|^2],\nonumber
\end{align}
where $\delta^2$ is the  variance of noise. For $\mathbb{E}[|I_{k,u}|^2]$,  let's first consider cases when $G(u)\neq G(k)$ and $G(u)\in\mathcal{K}_s$. Thus, $\hat{\mathbf{g}}_{s,k}$ and $\mathbf{v}_{s,u}$ are orthogonal to each other:
\begin{align}
\hat{\mathbf{g}}_{s,k}^\herm\mathbf{\Pi}(\mathbf{B}_{s,u})\hat{\mathbf{g}}_{s,u}=0.\nonumber
\end{align}
Therefore, we define $\mathcal{R}_{G(k)}$ as the set of BSs that uses observations from $G(k)$-th pilot code for zero-forcing, i.e., $\mathcal{R}_{G(k)}=\{s|G(k)\in\mathcal{K}_s\}$. Therefore, when $G(u)\neq G(k)$, we have
\begin{align}
&\mathbb{E}[|I_{k,u}|^2]\nonumber\\
=&\mathbb{E}[|\sum\limits_{s=1}^S\eta_{s,u}^{\frac{1}{2}}\hat{\mathbf{e}}_{s,k}^\herm\frac{\mathbf{\Pi}(\mathbf{B}_{s,u})\hat{\mathbf{g}}_{s,u}}{\|\mathbf{\Pi}(\mathbf{B}_{s,u})\hat{\mathbf{g}}_{s,u}\|}|^2+\nonumber\\
&|\sum\limits_{s\not\in\mathcal{R}_{G(k)}}\eta_{s,u}^{\frac{1}{2}}\hat{\mathbf{g}}_{s,k}^\herm\frac{\mathbf{\Pi}(\mathbf{B}_{s,u})\hat{\mathbf{g}}_{s,u}}{\|\mathbf{\Pi}(\mathbf{B}_{s,u})\hat{\mathbf{g}}_{s,u}\|}|^2]\nonumber\\
=&\sum\limits_{s=1}^S\eta_{s,u}\sigma_{e,k,s}^2+\mathbb{E}[|\sum\limits_{s\not\in\mathcal{R}_{G(k)}}\eta_{s,u}^{\frac{1}{2}}\hat{\mathbf{g}}_{s,k}^\herm\frac{\mathbf{\Pi}(\mathbf{B}_{s,u})\hat{\mathbf{g}}_{s,u}}{\|\mathbf{\Pi}(\mathbf{B}_{s,u})\hat{\mathbf{g}}_{s,u}\|}|^2].
\label{Eq.int2ndterm}
\end{align}
Let us simplify the second term in (\ref{Eq.int2ndterm}):
\begin{align}
&\mathbb{E}[|\sum\limits_{s\not\in\mathcal{R}_{G(k)}}\eta_{s,u}^{\frac{1}{2}}\hat{\mathbf{g}}_{s,k}^\herm\frac{\mathbf{\Pi}(\mathbf{B}_{s,u})\hat{\mathbf{g}}_{s,u}}{\|\mathbf{\Pi}(\mathbf{B}_{s,u})\hat{\mathbf{g}}_{s,u}\|}|^2]\nonumber\\
=&\sum\limits_{j\not\in\mathcal{R}_{G(k)}}\sum\limits_{i\not\in\mathcal{R}_{G(k)}}\eta_{j,u}^{\frac{1}{2}}\eta_{i,u}^{\frac{1}{2}}\mathbb{E}[\hat{\mathbf{g}}_{j,k}^\herm\frac{\bar{\mathbf{g}}_{j,u}}{\|\bar{\mathbf{g}}_{j,u}\|}\frac{\bar{\mathbf{g}}_{i,u}^\herm}{\|\bar{\mathbf{g}}_{i,u}\|}\hat{\mathbf{g}}_{i,k}]\nonumber\\
=&\sum\limits_{s\not\in\mathcal{R}_{G(k)}}\eta_{s,u}\mathbb{E}[\frac{\bar{\mathbf{g}}_{s,u}^\herm}{\|\bar{\mathbf{g}}_{s,u}\|}\hat{\mathbf{g}}_{s,k}\hat{\mathbf{g}}_{s,k}^\herm\frac{\bar{\mathbf{g}}_{s,u}}{\|\bar{\mathbf{g}}_{s,u}\|}]\nonumber\\
=&\sum\limits_{s\not\in\mathcal{R}_{G(k)}}\eta_{s,u}\mathbb{E}[\frac{\bar{\mathbf{g}}_{s,u}^\herm}{\|\bar{\mathbf{g}}_{s,u}\|}\mathbb{E}[\hat{\mathbf{g}}_{s,k}\hat{\mathbf{g}}_{s,k}^\herm|\bar{\mathbf{g}}_{s,u}]\frac{\bar{\mathbf{g}}_{s,u}}{\|\bar{\mathbf{g}}_{s,u}\|}]\nonumber\\
=&\sum\limits_{s\not\in\mathcal{R}_{G(k)}}\eta_{s,u}\gamma_{s,k}\mathbb{E}[\frac{\bar{\mathbf{g}}_{s,u}^\herm}{\|\bar{\mathbf{g}}_{s,u}\|}\frac{\bar{\mathbf{g}}_{s,u}}{\|\bar{\mathbf{g}}_{s,u}\|}]\nonumber\\
=&\sum\limits_{s\not\in\mathcal{R}_{G(k)}}\eta_{s,u}\gamma_{s,k}.
\label{Eq.int2ndterm2}
\end{align}
Substituting (\ref{Eq.int2ndterm2}) back to (\ref{Eq.int2ndterm}), we can get 
\begin{align}
\mathbb{E}[|I_{k,u}|^2]&=\sum\limits_{s=1}^S\eta_{s,u}\sigma_{e,k,s}^2+\sum\limits_{s\not\in\mathcal{R}_{G(k)}}\eta_{s,u}\gamma_{s,k}.
\label{Eq.int2ndterm3}
\end{align}

Then, let's consider the case when $G(u)=G(k)$:
\begin{align}
\mathbb{E}[|I_{k,u}|^2]&=\sum\limits_{s=1}^S\eta_{s,u}((g-K_s+1)\gamma_{s,k}+\sigma_{e,k,s}^2)\nonumber\\
&+\sum\limits_{j=1}^S\sum\limits_{i\neq j}^S \eta_{j,u}^{\frac{1}{2}}\eta_{i,u}^{\frac{1}{2}}\gamma_{j,k}^\frac{1}{2}\gamma_{i,k}^\frac{1}{2}\Omega(g,K_j)\Omega(g,K_i).
\label{Eq.2ndMomentInt_4}
\end{align}
To achieve (\ref{Eq.2ndMomentInt_4}) follows similar procedure to get (\ref{Eq.userfulSinal2nd}). Combining (\ref{Eq.2ndMomentInt_4}) and (\ref{Eq.int2ndterm3}), we can get the expression for the variance of interference plus noise:
\begin{align}
&\sigma_{z_k}^2=\rho_d\sum\limits_{u\neq k}\sum\limits_{s=1}^S\eta_{s,u}\sigma_{e,k,s}^2+\rho_d\sum\limits_{G(u)\neq G(k)}\sum\limits_{s\not\in\mathcal{R}_{G(k)}}\eta_{s,u}\gamma_{s,k}+\nonumber\\
&\rho_d\sum\limits_{u\neq k,G(u)=G(k)}[\sum\limits_{s=1}^S\eta_{s,u}(g-K_s+1)\gamma_{s,k}+\nonumber\\
&\sum\limits_{j=1}^S\sum\limits_{i\neq j}^S \eta_{j,u}^{\frac{1}{2}}\eta_{i,u}^{\frac{1}{2}}\gamma_{j,k}^\frac{1}{2}\gamma_{i,k}^\frac{1}{2}\Omega(g,K_j)\Omega(g,K_i)]+\delta^2.\nonumber
\end{align}
The final SINR expression of (\ref{Eq.rateBound_caire}) is exhibited in (\ref{Eq.SINR2})\footnote{Note that for (\ref{Eq.rateBound_caire}), we can obtain its close-form expression for arbitrary settings of $T_d$ by substituting expressions of $\mathbb{E}[|D_k|^2]$ and $\delta_{z_k}^2$. In Sec. \ref{simulations}, we assume $T_d$ is infinite for simplicity, so that we can strike out its second term and directly use (\ref{Eq.SINR2}) for evaluation}. 
\begin{figure*}[!ht]
\normalsize
\begin{align}
&\text{SINR}_{k,ZFBF,(\ref{Eq.rateBound_caire})}=\frac{\rho_d|\sum\limits_{s=1}^S\eta_{s,k}^{\frac{1}{2}}\gamma_{s,k}^{\frac{1}{2}}\Omega(g,K_s)|^2+\rho_d\sum\limits_{s=1}^S\eta_{s,k}\gamma_{s,k}(g-K_s+1-\Omega^2(g,K_s))+\rho_d\sum\limits_{s=1}^S\eta_{s,k}\sigma_{e,k,s}^2}{\splitfrac{\delta^2+\rho_d\sum\limits_{s\not\in\mathcal{R}_{G(k)}}\gamma_{s,k}+\rho_d\sum\limits_{u\neq k}\sum\limits_{s=1}^S\eta_{s,u}\sigma_{e,k,s}^2+}{\rho_d\sum\limits_{u\neq k,G(u)=G(k)}[\sum\limits_{s=1}^S\eta_{s,u}(g-K_s+1-\Omega^2(g,K_s))\gamma_{s,k}]+\rho_d\sum\limits_{u\neq k,G(u)=G(k)}|\sum\limits_{s=1}^S\eta_{s,u}^{\frac{1}{2}}\gamma_{s,k}^{\frac{1}{2}}\Omega(g,K_s)|^2}}.
\label{Eq.SINR2}\\
&\text{SINR}_{k,ZFBF,(\ref{Eq.rateBound_caire}),asy.}=\frac{\rho_d|\sum\limits_{s=1}^S\eta_{s,k}^{\frac{1}{2}}\sqrt{g-K_s}\gamma_{s,k}^{\frac{1}{2}}\big|^2}{\delta^2+\rho_d\sum\limits_{s\not\in\mathcal{R}_{G(k)}}\gamma_{s,k}+\rho_d\sum\limits_{s=1}^S\sigma_{e,k,s}^2+\rho_d\sum\limits_{u\neq k,G(u)=G(k)}|\sum\limits_{s=1}^S\eta_{s,u}^{\frac{1}{2}}\sqrt{g-K_s}\gamma_{s,k}^{\frac{1}{2}}|^2}.\label{Eq.SINRasy}
\end{align}  
\hrulefill
\vspace*{4pt}
\end{figure*}
To investigate asymptotic analysis at the massive MIMO regime, we introduce a scalar factor $n$ for a set of parameters $g,[K_s]$ and $[\eta_{s,k}]$. Therefore, the number of BS antennas is $ng$, the number of used pilot codes by $\text{BS}_s$ is $nK_s$, and the power coefficients are reduced by a factor of n, i.e., $[\frac{\eta_{s,k}}{n}]$. Therefore, let $n$ go to infinity, we can obtain the asymptotic result of (\ref{Eq.SINR2}) as (\ref{Eq.SINRasy}) shows by utilizing the following property of Gamma function
\begin{align}
\lim_{n\to\infty}\frac{\Gamma(n+\alpha)}{\Gamma(n)n^\alpha}=1.\nonumber
\end{align}


\section{$\Rm_k^{\rm flat}$ approximation }\label{Flat}

In this part, we aim to justify the assumption of IID channels within each sector in (\ref{piecewise-lambda}). IID channel assumption in each sector actually results in a piece-wise flat covariance matrix, $\Rmod{flat}$. Justification of this assumption will be done by rate performance comparisons between covariance matrices created by 3GPP \cite{3GPP} model and piece-wise flat covariance matrix. 

Covariance matrices of 3GPP models with finite length ULA, $\Rmod{3GPP}$'s, are not necessarily circulant. We first define a circulant approximation as follows: $\Rmod{}= \Fmb {\rm diag}\left({\rm eig}(\Rmod{3GPP})\right)  \Fmb^\herm$, where ${\rm eig}(\cdot)$ returns a vector consisting of all the eigen values. 

The piece-wise flat approximation of $\Rmod{}$ is given by $\Rm_k^{\rm flat} = \Fmb\Lambdam_k^{\rm flat}\Fmb^\herm$, where 
\begin{eqnarray}
\Lambdam_k^{\rm flat}&=&{\rm diag} \left(\barlam{1}{k}^{\rm 3GPP}\onev_g^\herm,\ldots,\barlam{S}{k}^{\rm 3GPP}\onev_g^\herm\right),\nonumber
\end{eqnarray} where $\onev_g$ is an all-ones vector of size $g\times 1$ and $\barlam{s}{k}^{\rm 3GPP}$'s are obtained according to (\ref{barlam}) from the eigenvalues of $\Rmod{3GPP}$.

In this comparison, we assume ideal CSI is available for all users. As we want this comparison to be general, not specific to our scheme, we also do not use spatial filters at the receiver. We consider both ZFBF and CBF.  All $L$ users are given equal power, that is, power $\rho_d/L$. Focusing on a single fading block $f$, the received signal at user $k\in \{1,\ldots,L\}$ is given by (\ref{dlmimo}), where $\xv^\transp(f)$ is the precoded vector signal of size $1\times \MB$ transmitted by the BS
\begin{equation}
\xv^\transp(f) =  \infosigv^\transp(f)  \Vmb,\nonumber
\end{equation}
where $\infosigv^\transp=\begin{bmatrix}  \infosig_1 &  \infosig_2 & \ldots &  \infosig_L \end{bmatrix}$ is the information bearing signal with $\infosig_k\sim\Cc\Nc(0,1)$, and 
\begin{equation}
\Vmb(f) = \begin{bmatrix}  \vvb_{1}(f) &  \vvb_{2}(f) & \ldots &  \vvb_{L}(f) \end{bmatrix}\nonumber\end{equation}
 denotes the $L\times \MB$ precoder at fading block $f$.

For ZFBF, $\vvb_{k}(f)$ is in the direction of the unit-norm vector that is zero-forced to all other user channel estimates. For CBF, $\vvb_{k}(f)$ is in the direction of the unit norm vector which is given by normalizing the conjugate of the user's channel estimate. Also $\|\vvb_{k}(f)\|^2 = 1/L$. We can re-write the received signal as a summation of signal and interference-noise as follows:
\begin{align}
r_k=\sqrt{\rho_d}\mathbf{h}_k(f)^\herm\mathbf{v}_{k}(f)u_k+\sqrt{\rho_d}\sum_{k'\neq k}\mathbf{h}_k(f)^\herm\mathbf{v}_{k'}(f)u_{k'}+n_k.\nonumber
\end{align}

\begin{figure}
\centering
\includegraphics[width=3.5in]{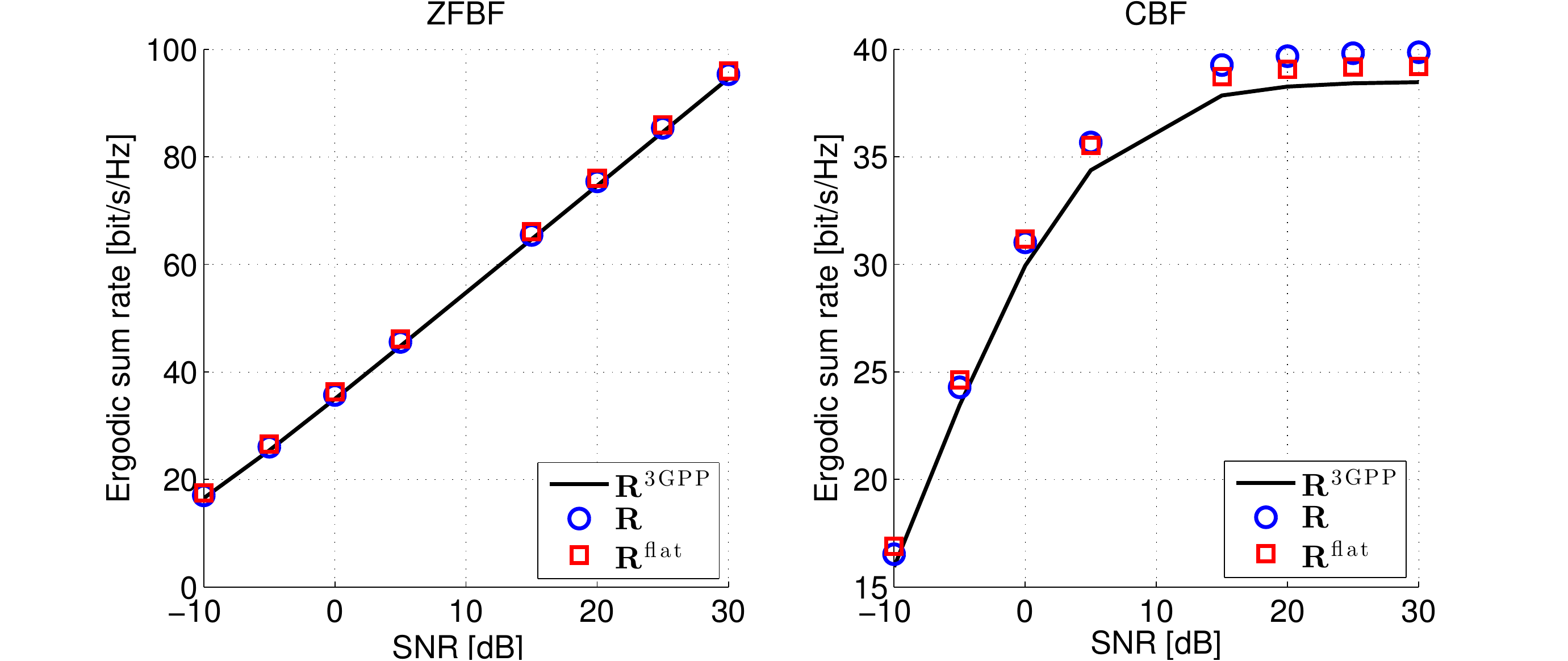}
\caption{Ergodic sum rate vs. SNR: for $[\mathbf{R}^{\text{flat}}]$, we have $S=10$, and $g=\frac{\MB}{S}=40$}
\label{Fig:sumRate_piece}
\vspace{-0.4cm}
\end{figure}

\begin{figure}
\centering
\includegraphics[width=3.5in]{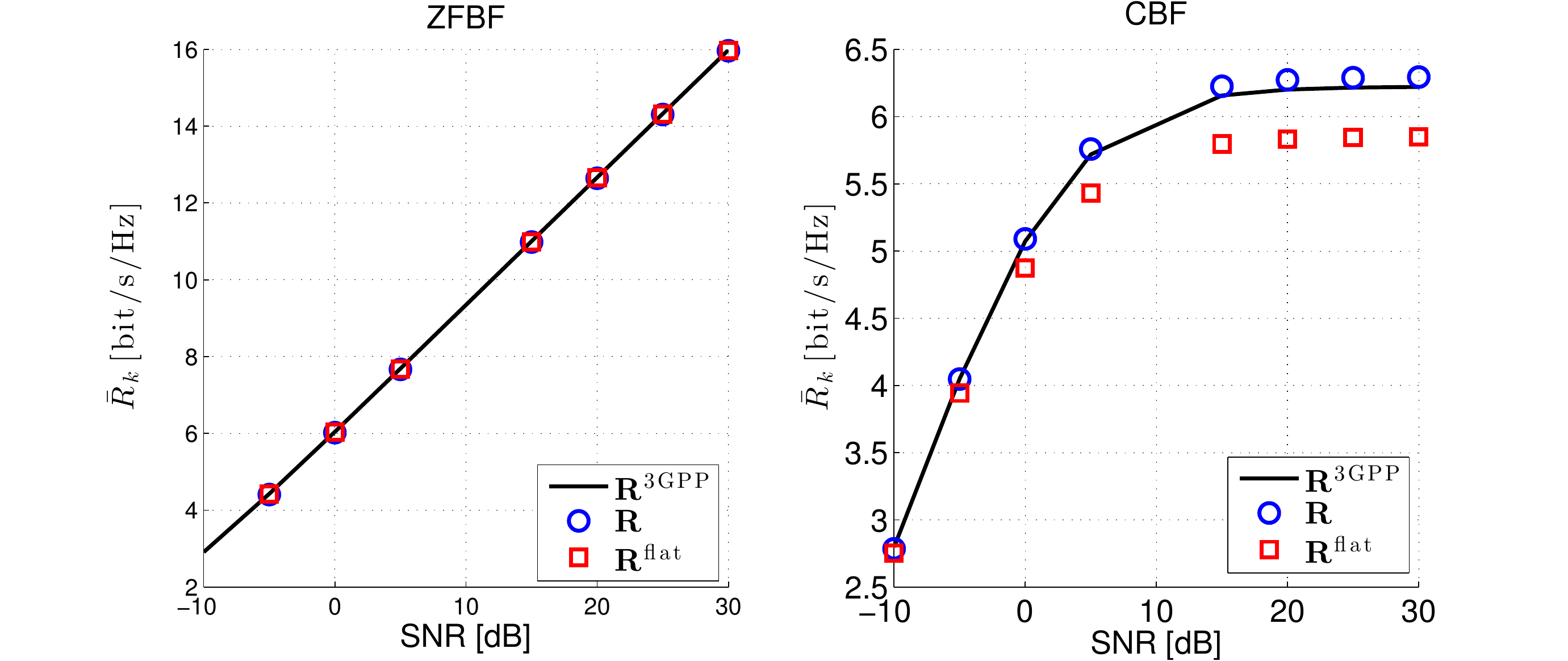}
\caption{Ergodic rate of individual UE vs. SNR: for $[\mathbf{R}^{\text{flat}}]$, we have $S=10$, and $g=\frac{\MB}{S}=40$}
\label{Fig:indRate_piece}
\vspace{-0.4cm}
\end{figure}

The ergodic rate of user $k$ is given by
\begin{align} 
\bar{R}_k&=\mathbb{E}\left[\log{(1+\frac{\rho_d|\mathbf{h}_k^\herm(f)\mathbf{v}_{k}(f)|^2}{\rho_d\sum_{k'\neq k}|\mathbf{h}_k^\herm(f)\mathbf{v}_{k'}(f)|^2+1})}\right],\label{barRate}
\end{align}
where the expectation is to average out the randomness of $\mathbf{h}_k(f)$. In this section Monte Carlo simulations are used to evaluate (\ref{barRate}) in  Fig. \ref{Fig:sumRate_piece}, \ref{Fig:indRate_piece}.
 
Consider a single cell with $\MB=400$ and $\MU =1$, we generate synthetic channel profiles for $3$ UEs with line-of-sight (LOS) and $3$ UEs without LOS, respectively. The large scale loss (pathloss plus shadowing) is neglected through simulations, so that we can better investigate the impact of channel covariance approximation. 
Simulating $[\mathbf{h}_k]$ by original channel covariance $[\Rmod{3GPP}]$,\footnote{Details of channel covariance extraction can be found in \cite{adhikary-mmWave-jsac}.} circulant approximation $[\mathbf{R}_k]$, and piecewise constant approximation $[\Rmod{flat}]$, respectively,  we make comparisons over ergodic rates for both ZFBF and CBF as Fig. \ref{Fig:sumRate_piece}, \ref{Fig:indRate_piece} show. 

Fig. \ref{Fig:sumRate_piece} exhibits comparisons of ergodic sum rates of $6$ UEs, while Fig. \ref{Fig:indRate_piece} shows the individual rate of one of UEs. For ZFBF, we can observe that in both figures the piecewise constant approximation is consistent with the original channel covariance in the sense of ergodic rate. Nevertheless, CBF is more sensitive to the piecewise constant approximation, since it may span the angular spectrum of interference signal and exaggerate its impact.

\vspace{1cm}
 \bibliographystyle{IEEEtran}
\bibliography{IEEEabrv,refs}
\end{document}